\documentclass[structuredabstract]{aa}  
\usepackage{graphicx}
\usepackage{ulem}

\usepackage{natbib}
\usepackage{multirow}
\usepackage{amsmath}
\usepackage{txfonts}

\usepackage[usenames]{color}
\usepackage{enumerate}
\usepackage{booktabs}
\usepackage[colorlinks,
	linkcolor = blue, 
	citecolor = blue,
	breaklinks = true]{hyperref}
	\usepackage{memhfixc}

\begin{document}

   \title{Evaporating Very Small Grains  as tracers of the UV radiation field in Photo-dissociation Regions
 }
   \author{
P. Pilleri\inst{1,2,3,4}		
\and 
J.  Montillaud\inst{1,2}        
\and
O. Bern\'e\inst{1,2}       
\and
C. Joblin\inst{1,2}
}
% (1) UniversitŽ de Toulouse; UPS-OMP; IRAP;  Toulouse, France
%(2) CNRS; IRAP; 9 Av. colonel Roche, BP 44346, F-31028 Toulouse cedex 4, France
\institute{Universit\'e de Toulouse; UPS-OMP; IRAP;  Toulouse, France
\and
CNRS; IRAP; 9 Av. colonel Roche, BP 44346, F-31028 Toulouse cedex 4, France
\and
Centro de Astrobiolog\'{\i}a (INTA-CSIC),
             Ctra. M-108, km.~4, E-28850 Torrej\'on de Ardoz, Spain
\and
%2
Observatorio Astron\'omico Nacional, Apdo. 112, E-28803 Alcal\'a de Henares, Spain			
}
\date{Received ????; Accepted????}
\offprints{P.~Pilleri, \email{p.pilleri@oan.es}} 
   \abstract
   {In photo-dissociation regions (PDRs), polycyclic aromatic hydrocarbons (PAHs)  may be produced by evaporation of  very small grains (VSGs) by the impinging UV radiation field from a nearby star.}
   { We quantitatively investigate  the transition zone between evaporating very small grains (eVSGs) and PAHs  in several PDRs. }
   {We studied the relative contribution  of PAHs and eVSGs to the mid-IR emission in a wide range of excitation conditions. We  fitted the observed mid-IR emission of PDRs by using a set of template band emission spectra of PAHs, eVSGs, and  gas lines.  The  fitting tool PAHTAT (PAH Toulouse Astronomical Templates) is made available to the community as an IDL routine. From the results of the fit, we derived the fraction of carbon $f_{\rm eVSG}$ locked in eVSGs and compared it to the intensity of the local UV radiation field. }
   {We show a clear decrease of $f_{\rm eVSG}$  with increasing intensity of the local UV radiation field, which supports the scenario of photo-destruction of eVSGs. Conversely, this dependence can be used to quantify the intensity of the UV radiation field for different PDRs, including unresolved ones.  }
%   , 
%with values in the range $\sim 50$ and $\sim 5\times10^4$ in Habing units. }   % conclusions heading (optional), leave it empty if necessary 
   { PAHTAT can be used to trace the intensity of the local UV radiation field in 
regions where eVSGs evaporate, which correspond to relatively dense ($n_H=[100,10^5]$\,cm$^{-3}$) and UV irradiated  PDRs  ($G_0=[100,5\times10^4]$) where H$_2$ emits in rotational lines. }
%Thanks to their improved spatial resolution, the forthcoming missions JWST and SPICA will allow us to apply our method for the study of PDRs in star-forming galaxies and at the surface of proto-planetary disks.

%   {Based on these results, we provide a tool to quantify the local UV radiation field in PDRs from the analysis of the mid-IR emission spectrum. We show that good results can be obtained both from resolved and unresolved objects.  Combined with the forthcoming missions such as JWST and SPICA, it will allow studying the spatial variations of the UV radiation field in more distant galaxies.} 

  
\titlerunning{Evaporating Very Small Grains as tracers of the UV field in PDRs}

   \maketitle
%
%________________________________________________________________
%==============================
% Introduction

\section{Introduction}	\label{sec:introduction}
The mid-infrared (mid-IR) spectrum of photo-dissociation regions (PDRs) is dominated by the aromatic infrared bands (AIBs) attributed to the emission of polycyclic aromatic hydrocarbons   \citep[PAHs, ][]{leger84, allamandola85, joblin11}.  Very small grains (VSGs) are another class of mid-IR emitters  if they are small enough to be stochastically heated \citep{desert90}. The VSGs are generally considered to be carbonaceous \citep{desert86} but \citet{li01a} concluded that the presence of very small silicate grains is not excluded and can involve up to 10\% of interstellar Si. 
\citet{cesarsky00a} suggested  a link between  PAHs and VSGs, in which the latter are processed by the UV radiation field or by shocks giving birth to free PAHs. 

Using blind-source separation methods, \citet{rapacioli05} and \citet{berne07a} extracted three independent spectra from the mid-IR emission, two molecular PAH-type spectra attributed to neutral and ionised PAH populations, respectively and one consisting of both continuum and broad band emissions, which was attributed to  some VSGs. The relative spatial distribution between PAHs and  these VSGs additionally supported the idea that  free flying PAH molecules  are produced by photo-evaporation of VSGs under UV irradiation at the border of PDRs \citep{rapacioli05,berne07a}.  Although they are called VSGs, this population is not strictly that included in the \citet{desert90} model: it is better described by the PAH grains of \citet{li01b}.
In the following, we will refer to them as evaporating VSG (eVSGs). 

Using the template spectra derived by  \citet{rapacioli05} and \citet{berne07a},  \citet{joblin08} and \citet{berne09b} were able to fit the mid-IR spectra of objects in different evolutionary stages, such as planetary nebulae and proto-planetary disks. 
A spectrum attributed to large ionised PAHs  (PAH$^x$) was introduced to fit highly irradiated objects.  
In \citet{berne09b} and \citet{berne09}, the relative intensities of the different mid-IR components were related to physical quantities such as the spectral type of the central illuminating star for proto-planetary disks, or the ionisation parameters and radiation field for the Monoceros R2 compact HII region. While it is clear, in all the above studies, that the evolution of eVSGs into PAHs is related to the intensity of the UV field, this has never been quantified.

In this paper, we search for a quantitative link between the intensity of the UV radiation field and the chemical evolution of the AIB carriers  in terms of processing of eVSGs into PAHs.  Section \ref{pdrsample} describes the PDR sample.  Section \ref{spectralfitting} presents the fitting tool we have developed to decompose mid-IR spectra into chemical populations, which takes into account dust extinction. 
We apply  this method to our PDR sample and present the results in Sect. \ref{application}. Section \ref{linking} describes the relation between the local physical conditions and the evolution of the AIB carriers 
and illustrates the potential of our method for characterising the UV field in a variety of PDR environments.

\section{The PDR sample}

\label{pdrsample}

\begin{table*}[htdp]
\begin{center}
\caption{Observational parameters for PDRs with spatially resolved mid-IR populations.
}	
\label{tab:pdrinput}
\begin{tabular}{c|ccccc|cc|c}

\toprule

\multirow{2}{*}{Object}	& 	\multirow{2}{*}{Star}	& 	Distance						&	Spectral 				&  	Kurucz	 			& Radius	$^*$	& 		$d_{\rm front}$&	$G_0^{\rm front}$	& Instrument \\
					&					&	$[$pc$]^{(a)}	$					&	Type					&	Spectrum $^*$ $[$K$]$	&$[R_{\sun}]$	&		$[\arcsec]$		&		 \\
\midrule
	\object{NGC~7023~NW}		&\multirow{3}{*}{HD 200775}&	\multirow{3}{*}{$430$}		&	\multirow{3}{*}{B3Ve --- B5$^{(b,c,d)}$}	& 	\multirow{3}{*}{$2 \times 15000^{(e)}$}	&\multirow{3}{*}{10}	&42	&	2600	 & Spitzer IRS-SL\\
	\object{NGC~7023~S}		&					&								&						&						&			&			55	&	1500	&ISO-CAM cvf	 		 \\
		\object{NGC~7023~E}		&					&								&						&						&			&			155	&	250& ISO-CAM cvf\\
\midrule
	\object{NGC~2023~N}		&\multirow{2}{*}{HD 37903}&	\multirow{2}{*}{470}&	\multirow{2}{*}{B1.5V$^{f}$}	& 	\multirow{2}{*}{24000}	&\multirow{2}{*}{6}			&		164	&	400		& Spitzer IRS-SL\\
	\object{NGC~2023~S}		&					&								&						&						&				&		67	&	4000& Spitzer IRS-SL \\
\midrule
	\object{$\rho$-Oph} filament			& HD 147889		&	118					&	B2III --- B2V			&	22000			&	5			&		610  &	520& ISO-CAM cvf\\
\bottomrule
\multicolumn{9}{l}{$^{*}$ Derived from spectral type.}\\
\multicolumn{9}{l}{ (a) \citet{van-den-ancker97} 	---	(b) \citet{racine68}	--- (c) \citet{finkenzeller85} ---  (d) \citet{witt06}	--- (e) \citet{alecian08}  }\\
\multicolumn{9}{l}{  (f) \citet{diplas94}}\\
\end{tabular}
\end{center}
\end{table*}%

\begin{table*}[thdp]
\begin{center}
\caption{The  {\it Spitzer} IRS observational parameters  and the assumed physical parameters for PDRs with mixed mid-IR populations. }
\label{tab:otherpdrs}
\begin{tabular}{c|cc|ccc}
\toprule
\multirow{2}{*}{Object}	&	Position						& Aperture						&	 	$G_0$			&		$n_{\rm H}$				&	Ref.	\\
					&	$[\alpha_{2000}, \delta_{2000}]$	&								&					&		[cm$^{-3}$]		&			\\			
\midrule
\object{Horsehead}			&	(05:40:53.8; -02:28:00)			& $9\arcsec \times 7\arcsec$				&		100				&		$2\times10^5$		&	\citet{habart05}	\\
\object{Ced 201}				&	(22:13:25; 70:15:03)				& $22\arcsec\times22\arcsec$			&		300				&		$4\times10^2$		&\citet{young-owl02}	\\
\object{IC 63}				&	(00:59:01; 60:53:19)				& $30\arcsec\times40\arcsec$			&		1100				&		$1\times10^5$		&\citet{gerin03}	\\
\object{Parsamyan 18 S}		&	(06:59:41; -07:46:45)			&$9\arcsec\times8\arcsec$				&		3500				&		$1\times10^4$		&\citet{ryder98} \\
\object{Parsamyan 18 N}		&	(06:59:41; -07:46:12)			& $4\arcsec\times8\arcsec$				&		5000				&		$1\times10^4$		&\citet{ryder98} \\
\object{Orion Bar}				&	(05:35:21.4; -05:25:15)			& $4\arcsec\times8\arcsec$				&		$4\times10^{4}$	&		$2\times10^5$		&\citet{tauber94}		\\	
%Monoceros R2 (P1)		&	(06:07:45; -06:23:20)			& $4\arcsec\times4\arcsec$			&	0.38			&	$1.6\times10^{5}$	&		\citet{berne09}	\\	
%Monoceros R2 (P3)		&	(06:07:47; -06:22:59)			& $4\arcsec\times4\arcsec$			&	0.32			&	$3.7\times10^{4}$	&		\citet{berne09}	\\
\bottomrule
\end{tabular}
\end{center}
\end{table*}

We analysed mid-IR observations of several PDRs that span a wide range of UV irradiation conditions. The majority of the observations discussed here were obtained with the short-low (SL) module of the Infrared Spectrograph (IRS) \citep{houck04} onboard {\it Spitzer} \citep{werner04} in spectral mapping mode. Data reduction was performed with the CUBISM software \citep{smith07b} and consisted of cube assembling, calibration, flux correction for extended sources and bad pixel removal. For the objects that were not observed with IRS, we used ISOCAM data available in the ISO data archive as highly processed data products \citep{boulanger05}.

\subsection{PDRs with spatially resolved mid-IR populations}

 Our sample contains several PDRs (Table \ref{tab:pdrinput})  where the emission from the PAH and eVSG populations (see Sect. \ref{spectralfitting}) can be spatially isolated.
These PDRs are referred to as {\it PDRs with spatially resolved mid-IR populations}.
It is possible to study the variations of the mid-IR spectral properties in these sources (Sect. \ref{spectralfitting}) and link them to the local physical properties such as the local density and UV field (Sect. \ref{linking}) as a function of the depth inside the PDR.
To estimate the intensity of the UV radiation field at the PDR front, we used modelled stellar spectra from \citet{kurucz93} and applied a dilution factor based on their projected distance to the PDR front. All parameters used to calculate the intensity of the UV radiation field at the PDR fronts are reported in Table \ref{tab:pdrinput}.  In the following, we will express the
UV field intensity $G_0$ in terms of the Habing field, which corresponds to $1.6 \times 10^{-3}$ erg cm$^{-2}$ s$^{-1}$ when integrated  between 91.2 and 240 nm \citep{habing68}.
The geometrical model assumed for these PDRs is discussed in Appendix \ref{subsec:profiles}.

\paragraph{$\rho-$oph filament} Assuming typical spectral properties for the B2 star HD 147889 and using geometrical dilution to the projected distance of 0.4\,pc between the star and the PDR leads to an intensity of the UV radiation field of $G_0 \sim 520$. 

\paragraph{NGC 2023} 
For NGC 2023 an extinction of A$_V$ = 1.1 mag was applied, reflecting  absorbing dust around the illuminating star  HD 37903 \citep{compiegne08}. 
Using a distance of 0.38 and 0.13 pc for the north and south PDRs, respectively yields an estimate of the UV radiation field intensity of $G_0 \sim 400$ and  $G_0 \sim 4000$. These estimates are consistent with previous values obtained in the far-IR study by \citet{burton98}.

\paragraph{NGC 7023} The estimate of the UV radiation field intensity at the PDR front in \object{NGC~7023~NW} is more uncertain.  \citet{chokshi88} estimated  $G_0\sim2.4\times10^3$  through far-IR observations and modelling, but values as high as $G_0\sim10^4$  have been proposed to explain the observed H$_2$ ortho-to-para ratio \citep{fuente99} towards this PDR. This value agrees with the value determined by assuming typical stellar properties for the corresponding spectral type of the illuminating binary star \citep{alecian08} and geometrical dilution to the PDR front. However, extinction between the star surface and the PDR may attenuate the radiation field, yielding a lower value at the PDR front. In this work, an extinction correction of $A_{\rm V}$ = 1.5 was derived from the IUE spectrum measured on the star and was applied on the corresponding Kurucz spectrum. Assuming that the same extinction factor applies between the star and the PDR leads to an estimate for the NW PDR of $G_0  \sim 2.6\times10^3$, which is consistent with the value of \citet{chokshi88}.

\subsection{PDRs with mixed mid-IR populations}

 In the previous section we have defined  the PDRs with spatially resolved mid-IR populations as those where
the PAHs and eVSGs can be spatially isolated. The PDRs of our sample that do not follow this
criterion are referred to as {\it PDRs with mixed mid-IR populations}. The fact that  in these sources the emission from PAHs and eVSGs 
cannot be isolated may result from different effects, e.g. from the small spatial extent
of the PDR, a lower signal-to-noise ratio in the data, specific geometrical properties, etc. For each of these sources we  analysed an averaged spectrum towards the peak of the mid-IR 
emission. 
The sample of PDRs with mixed mid-IR populations comprises a
wide range of irradiation conditions, from $G_0  \sim 100 $ for the Horsehead Nebula to $G_0  \sim 4\times10^4$ for the Orion Bar.
For these well-studied objects, a reliable estimate of the intensity of the UV radiation field
can be found in the literature.

\section{A spectral fitting tool for studying the chemical evolution of AIB carriers: PAHTAT} 
\label{spectralfitting}

Our motivation here is to provide a tool that allows us to decompose observed mid-IR  (5.5-14\,$\mu$m) spectra into a limited set of components
that can provide information on the chemical nature of the emitters.
We note that there already exists an AIB fitting tool called PAHFIT, proposed by \citet{smith07}, which decomposes mid-IR spectra
using a collection of lorentzians and black body functions with a large number of parameters. While PAHFIT provides high-quality fits
that are useful e.g. to extract the integrated intensity from individual bands, our aim is to decompose the observations using 
band template spectra for PAHs and eVSGs,  minor additional unattributed PAH features, gas lines, and a function to describe the continuum.  The corresponding IDL code PAHTAT (PAH Toulouse Astronomical Templates) is available for download\footnote{\url{http://userpages.irap.omp.eu/~cjoblin/PAHTAT}.}.

\subsection{Band emission}

\begin{figure}[ht]	
\centering
\includegraphics[width =0.92\hsize ]{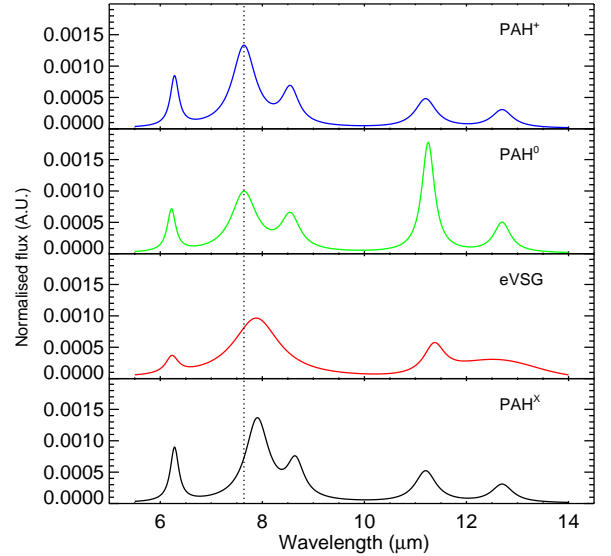}
\caption{Template spectra used in the fitting procedure. The eVSG spectrum from \citet{joblin08} was slightly improved to better fit the observed spectrum by adding a plateau in the 12-14\,$\mu$m range. Spectra were normalised to their integrated intensity between  5.5 and 14 $\mu$m.}
\label{fig:templates}
\end{figure}

 Applying  blind-signal separation methods to ISO and Spitzer spectro-imagery data of several PDRs \citet{rapacioli05} and \citet{berne07a}  extracted from the observations a  set of three elementary mid-IR spectra that can reproduce the AIB emission.  \citet{joblin08} and \citet{berne09b} found that a fourth spectral component was needed to explain the observed spectra in highly irradiated environments.
PAHTAT is based on these four   astronomical template band spectra (Fig. \ref{fig:templates}).
Here, we briefly present the spectral properties of this set of templates and their chemical assignment.
 Details on the spectral characteristics of the templates (e.g. relative band strength, position, width etc.), 
how they were obtained as well as the populations they represent can be found in \citet{joblin08}. 
These templates can be downloaded on the PAHTAT webpage.
The first component is the neutral PAH (PAH$^0$) spectrum, which is mainly characterised by a high C-H (11.2 $\mu$m) over C-C (7.7 $\mu$m) band ratio.  The second component is the PAH cation (PAH$^+$) spectrum, which displays on the contrary a low 11.2 $\mu$m over 7.7 $\mu$m band ratio.
The third  PAH  component is associated to  large ($\sim$ 100 C atoms) ionised PAHs, referred to as PAH$^{x}$, whereas the PAH$^+$
and PAH$^0$ populations would be dominated by smaller species.
Its spectrum shares the same characteristics as that of PAH$^{+}$ but has a ``7.7'' $\mu$m band that is shifted by 0.15 $\mu$m to the red compared to PAH$^{+}$ \citep[for a discussion of the reason for this shift, see][]{joblin08}. 
 The fourth component is the eVSG spectrum, whose carriers are still discussed and  have typical sizes of $\sim 500$ C atoms \citep{rapacioli05}.
 PAH clusters \citep{rapacioli06} or complexes of PAHs with Fe atoms \citep{simon09} have been proposed.
 eVSGs are thought to  carry continuum emission, but here we have treated band and continuum emission separately.  The eVSG band spectrum is made of  broad features at 7.7 and 11.2 $\mu$m and  has no 8.6 $\mu$m band (Fig. \ref{fig:templates}).
We improved this spectrum as compared to \citet{joblin08} by adding a  Gaussian plateau centred at 12.6\,$\mu$m with a FWHM of 1.76\,$\mu$m and a relative intensity compared 
to the 7.7\,$\mu$m band of 0.36. This plateau was introduced to better fit the extracted VSG component of \citet{rapacioli05} and \citet{berne07a}. 
The continuum model is described in the following section.

%\subsubsection*{The eVSG band spectrum}

\subsection{The continuum}
\label{sectcont}
In \citet{rapacioli05} and \citet{berne07a} it was shown that some continuum emission is spatially associated with the band emission of eVSGs.
 In mild UV field PDRs, the continuum emission is observed to be linear up to $\sim20 \mu$m \citep{berne07a, compiegne08}. In high UV field PDRs,
such as the Orion Bar, the shape of the continuum is observed to change due to a contribution from big grains at thermal equilibrium. For most of our
sources  we found that the 5.5-14 $\mu$m continuum can be fitted by a single slope. For high UV field PDRs a bilinear continuum is used.
The choice of the continuum (single slope or bilinear) can be set by the user in PAHTAT.

\subsection{Minor PAH features}
Three additional features are present in the mid-IR spectra of most PDRs, at  6.7, 11.55 and 13.5 $\mu$m. These features are best represented by three lorentzians at 6.7, 11.55 and 13.5\,$\mu$m with a FWHM of 0.2, 0.4 and 0.25\,$\mu$m, respectively. These features also appear in the PAH$^+$ and PAH$^0$ components of \citet{rapacioli05}. However, they are much fainter than the principal PAH features and were not included in the band template spectra. 
Their intensities are left as free parameters in the fit. However, these features are minor and their introduction does not change the results significantly. 

\subsection{Gas lines}
Several gas lines fall in the 5.5-14\,$\mu$m range, such as the  H$_2$ S(2) to S(5) rotational lines, and fine structure lines from ionised gas such as [ArII], [NeII] and [SIV]. Some of these lines may be blended with PAH/eVSG features at the spectral resolution of ISOCAM-cvf and IRS-SL. Therefore, these features cannot be simply subtracted but they rather need to be treated simultaneously with the PAH and eVSG bands. In PAHTAT, we assumed Gaussian profiles with a fixed central wavelength and a FWHM given by the average spectral resolution R of the instrument (R = 45 for ISOCAM, and R = 80 for IRS).

\subsection{Extinction correction}
Finally, our model takes into account  extinction  along the line of sight by dust within the PDR, which is caused mainly by  silicate grains. Assuming that the emitting and absorbing materials in the considered column of material  are fully mixed\footnote{PAHTAT gives the possibility to also use a foreground extinction correction $e^{-\tau_\lambda}$, although we assumed a mixed extinction model here.}
 \citep[see, for instance][]{disney89, smith07} leads to a correction factor $(1-e^{-\tau_{\lambda}})/{\tau_{\lambda}}$. The optical depth $\tau_\lambda$ at each wavelength is calculated using the relation
\begin{equation}
\tau_\lambda = C_{\rm ext}(\lambda) \times N_{\rm H},
\end{equation}
\noindent where $C_{\rm ext}(\lambda)$ is the extinction cross-section per H nucleon, calculated by 
\citet{weingartner01} for $R_{\rm V} = 5.5$.

\begin{figure}[t]	
\centering
\includegraphics[width =0.92\hsize ]{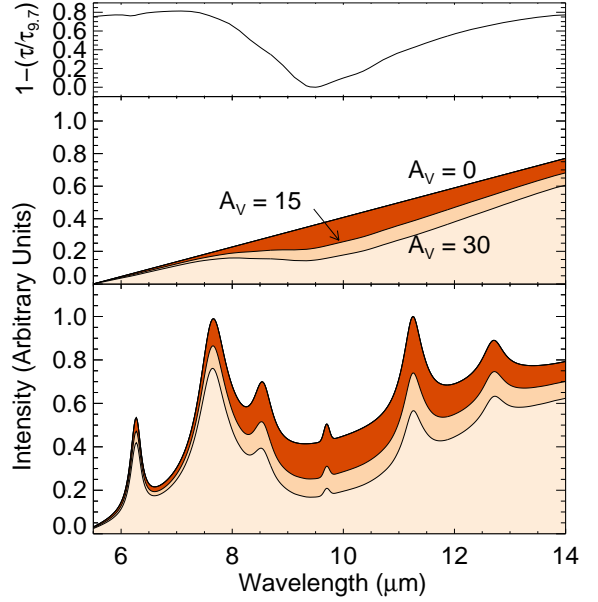}
\caption{Effect of the extinction correction on a linear continuum and on a synthetic PDR spectrum composed of continuum, band emission, and gas lines. }
\label{fig:extcorrec}
\end{figure}

\subsection{Summary of the model}
In summary, the modelled mid-IR emission $I_{\rm model}$ is written as
\begin{equation}
 I_{\rm model}  =  \left( I_{\rm bands}  + I_{\rm continuum}+ I_{\rm gas} + I_{\rm mbands}   \right) \frac{1-e^{-C_{\rm ext}~N_{\rm H} } }{C_{\rm ext}~N_{\rm H}}.
 \end{equation}
The adjusted parameters are $I_{\rm bands}$, which represents the intensity of the band template spectra  of PAH$^{0,+,x}$ and eVSGs, $I_{\rm continuum}$, which is the (bi-)linear continuum emission, $I_{\rm gas}$, which consists of  the entire mid-IR H$_2$ and ionised gas lines, and $I_{\rm mbands}$ to account for the minor band features at 6.7, 11.5 and 13.5\,$\mu$m. The hydrogen column density $N_{\rm H}$ is left as a free parameter in the fit. Hereafter, $N_{\rm H}$ will be expressed, for simplicity, in magnitudes of visual extinction  along the line of sight, defined as $A_{\rm V} =  N_{\rm H}  / (1.8 \times 10^{21}$ cm$^{-2}$) mag. We tested the effect of using different  extinction curves, corresponding to different values of $R_{\rm V}$, and found a difference of up to 20\% in the estimated column density but no significant effect on the population weights when using $R_{\rm V} = 3.1$ instead of $R_{\rm V} = 5.5$. The extinction correction can significantly modify the shape and intensity of the mid-IR spectrum. As an example, Fig. \ref{fig:extcorrec} shows the effect of the extinction correction to a linear continuum and a synthetic spectrum for different values of $A_{\rm V}$. This shows an example of how different (i.e. non-linear) continua can be obtained using just a linear continuum and an extinction correction.  On the other hand,  to obtain significant information on the column density, this must be high enough to modify the spectrum significantly. This happens at an $A_{\rm V} \gtrsim 3$, which we defined as our threshold. 

\subsection{$I_0$, the emission radiated by AIB carriers}
\label{section_contcorr}

\begin{figure}[ht]	  
\centering
\includegraphics[width =0.98\hsize, trim = 0cm 0cm 0cm 0cm]{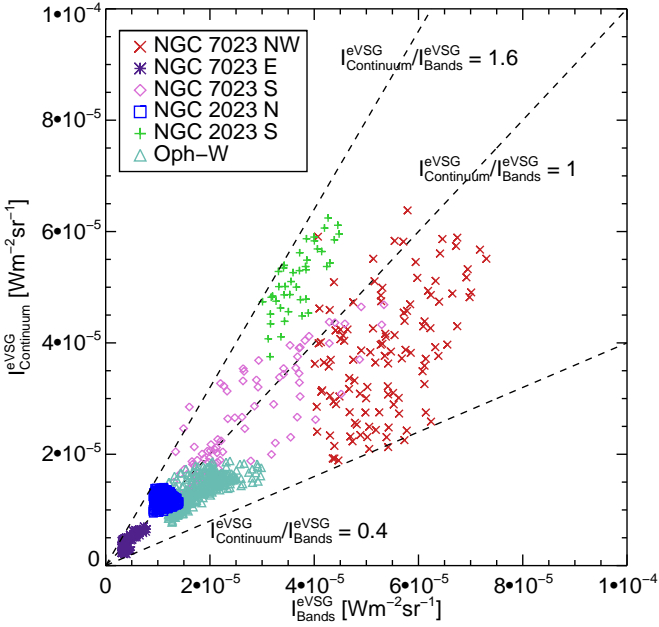}
\caption{Pixel-to-pixel correlation of the two components  $I^{\rm eVSG}_{\rm bands}$ and $I^{\rm eVSG}_{\rm continuum}$  for the resolved PDRs of Table \ref{tab:pdrinput}. For each PDR, a mask  was applied to show only the brightest pixels in $I_{\rm eVSG}^{\rm bands}$, i.e. the pixels towards the PDR front (see also Fig. \ref{fig:pdrprofiles}).  }
\label{fig:correlationcont}
\end{figure}

The total emission radiated by AIB carriers in both aromatic bands and mid-IR continuum of eVSGs (cf. Sect. \ref{sectcont}) is given by
\begin{equation}
 I_{0} =  I_{\rm bands} + I_{\rm continuum}^{\rm eVSG} = I^{\rm PAH}_{\rm bands}+ I_{\rm bands}^{\rm eVSG}+ I_{\rm continuum}^{\rm eVSG},
\end{equation}
where $ I^{\rm PAH}_{\rm bands}$ is the intensity of all the PAH populations (PAH$^+$, PAH$^0$ and PAH$^x$). $I_{\rm bands}^{\rm eVSG}$ and $I_{\rm continuum}^{\rm eVSG}$ are the contribution of eVSGs to the bands and the continuum, respectively. While $ I^{\rm eVSG}_{\rm bands}$ is a direct output of PAHTAT, 
$I^{\rm eVSG}_{\rm continuum}$ has to be computed separately. Considering that in moderate UV field regions ($G_0 \lesssim 4000$)
 emission from  big grains is negligible at $\lambda \lesssim 14$\,$\mu$m, we  considered that for these PDRs  $I^{\rm eVSG}_{\rm continuum}=I_{\rm continuum}$.
Hence using the data in this region, we can calibrate a relation between  $I^{\rm eVSG}_{\rm continuum}$ and  $I^{\rm eVSG}_{\rm bands}$. Figure \ref{fig:correlationcont} shows the 
pixel-to-pixel correlation of the two extracted components $I^{\rm eVSG}_{\rm bands}$ and $I_{\rm continuum}$  for all  resolved PDRs (where $G_0 \lesssim 4000$).
The dispersion of this correlation  suggests that the band and continuum emissions are not strictly related and that the band-to-continuum ratio can be affected by factors that still have to be understood, such as chemical composition and size distribution. 
To quantify the continuum intensity we  therefore used 
the limits given by the cloud of points in Fig.\,\ref{fig:correlationcont}, which gives 
$I^{\rm eVSG}_{\rm continuum} = (1\pm 0.6) \times I^{\rm eVSG}_{\rm bands}$. 
The contribution of eVSGs to emission at longer wavelengths ($\lambda \gtrsim 14\,\mu$m) is neglected based on the results by \citet{draine07}, who showed that it   is not significant for PAH grains of size $\lesssim 10$\AA\, ($\sim 1200$ C atoms), which includes the size of eVSGs \citep{rapacioli05}.

After defining the total eVSG emission, we can use in the following the ratio of this value to the total AIB emission as a tracer of the evolution of eVSGs in different environments: 
\begin{equation}
\label{fevsg}
 f_{\rm eVSG}= \frac{I^{\rm eVSG}_{\rm bands}+I^{\rm eVSG}_{\rm continuum}}{I_0} =  (2\pm0.6)\times \frac{I^{\rm eVSG}_{\rm bands}}{ I_0},
\end{equation} 
\noindent which can be derived   directly  from PAHTAT.

\begin{figure*}[t]	  
\centering
\includegraphics[width =0.95\hsize]{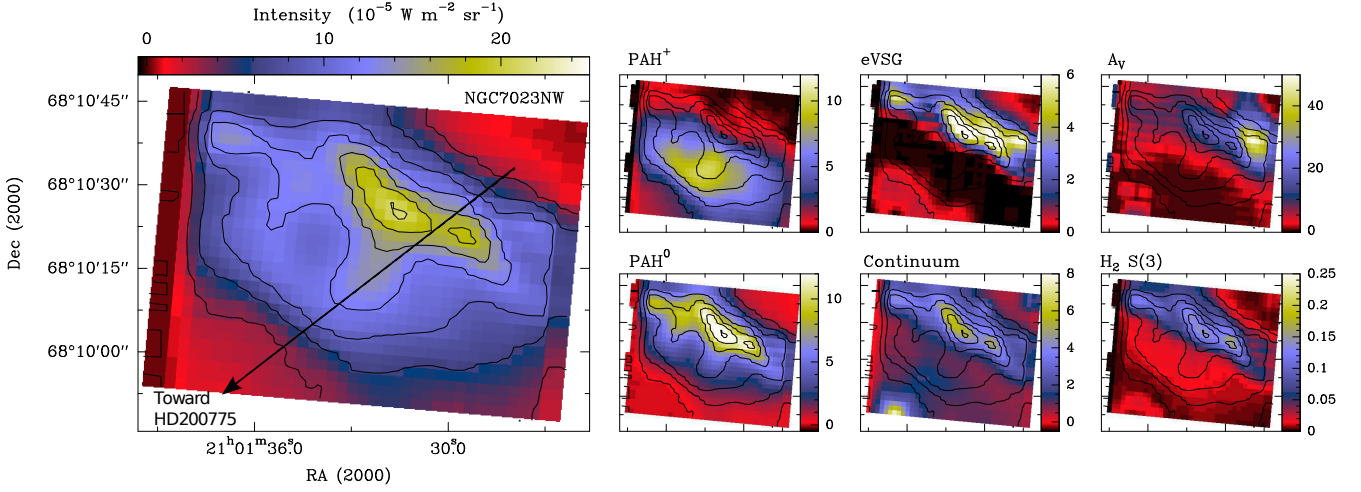}
\caption{Maps of the different components obtained using PAHTAT to fit the observed mid-IR spectra of the \object{NGC~7023~NW} IRS spectral cube.
Integrated intensities  in the 5.5-14 $\mu$m range of the total emission (left map) and of the different 
band template components (central panels) are displayed. The map of the column density expressed in $A_{\rm V}$ is also shown, as well as the flux in the H$_2$ S(3) line (right maps). 
 The black arrow shows the direction of the spatial cut studied in this work (Fig.~\ref{fig:pdrprofiles}).
Contours  of the total emission map are given in steps of $3\times 10^{-5}$\,W\,m$^{-2}$\,sr$^{-1}$ and are reported for reference in all the maps.  The color intensity scales are in units of $10^{-5}$\,W\,m$^{-2}$\,sr$^{-1}$, while the column density is expressed in magnitude of visual extinction. Results for the other resolved PDRs  can be found in Fig. \ref{fig:allmaps}.}
\label{fig:templatesmap}
\end{figure*}

\begin{figure*}[ht]	
\centering
\includegraphics[width =0.49 \hsize]{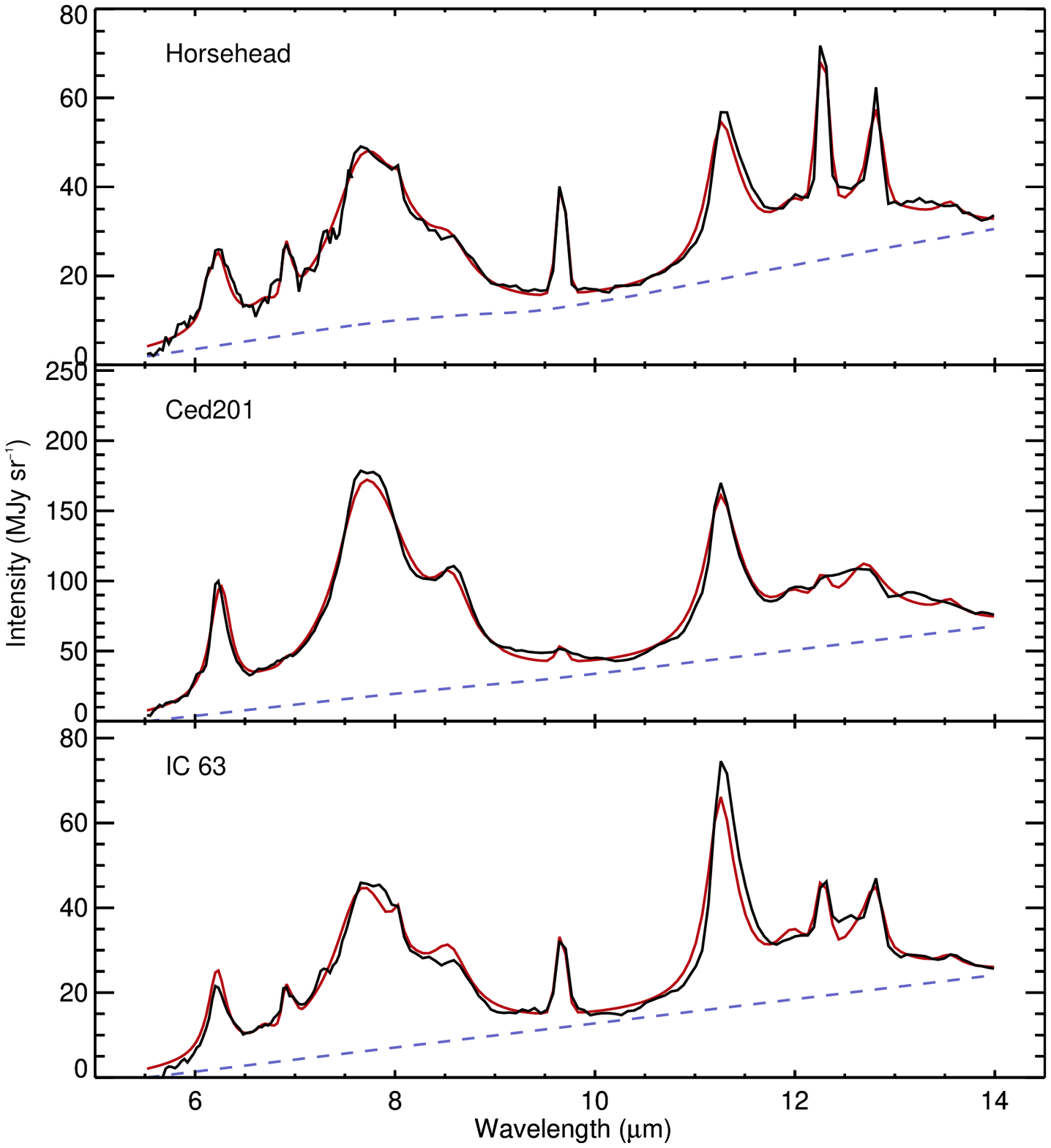}
\includegraphics[width =0.49 \hsize]{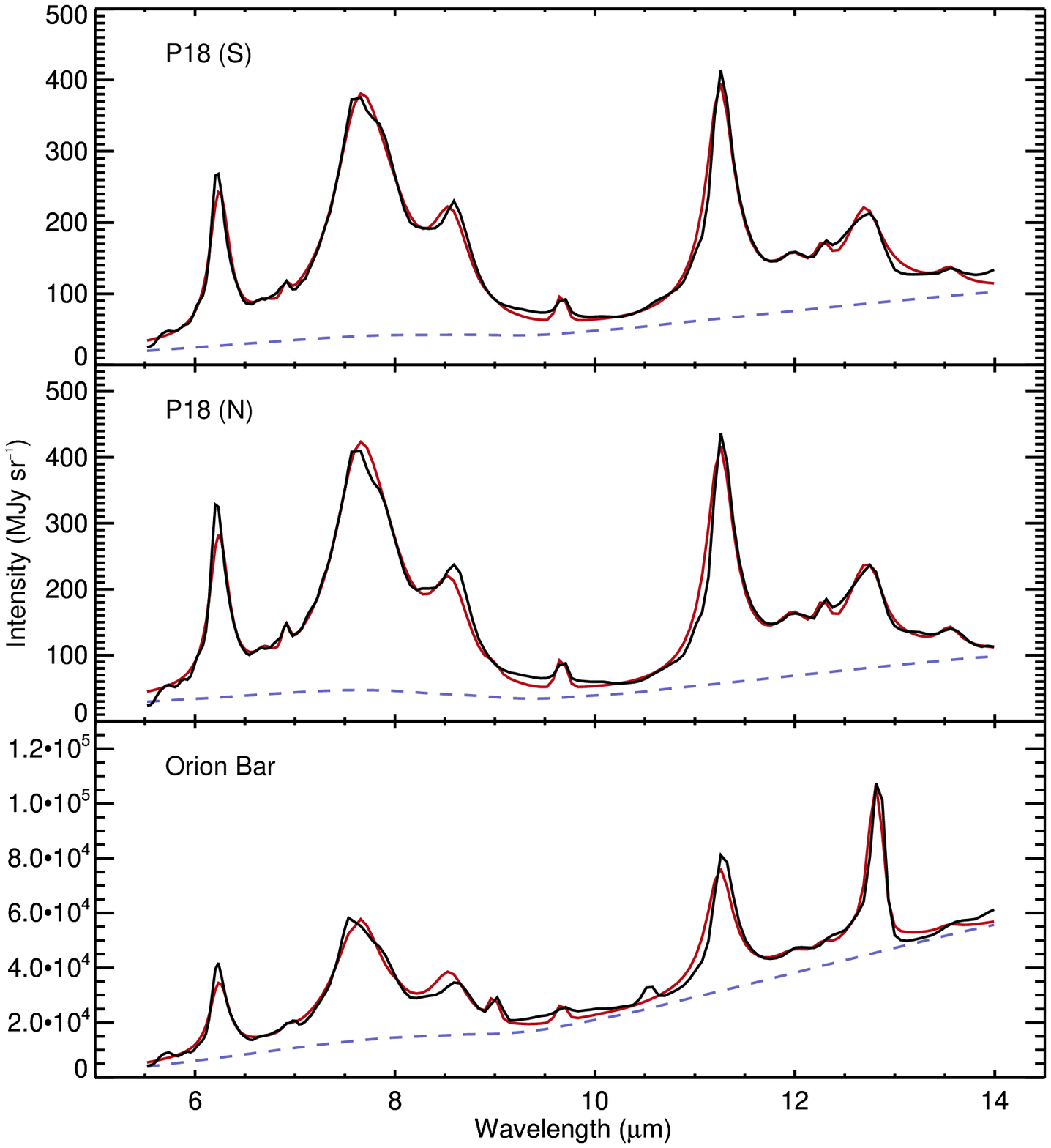}
\caption{ Mid-IR spectra (black) obtained with IRS and their fit with PAHTAT (red) for the PDRs listed in Table \ref{tab:otherpdrs}. All quantities derived from the fit are reported in Table
\ref{table_unresolvedresults}. The dashed line represents the continuum determined by the fit, corrected for the extinction along the line of sight. }
\label{fig:spectra_aib}
\end{figure*}

\section{Results of PAHTAT}
\label{application}

We used PAHTAT to fit the emission spectra of the PDRs with spatially resolved (Table \ref{tab:pdrinput}) and mixed (Table \ref{tab:otherpdrs})  mid-IR populations.
 
\subsection{PDRs with spatially resolved mid-IR populations}

We applied PAHTAT for each spatial pixel of the spectro-imagery cubes of the resolved PDRs listed in Table \ref{tab:pdrinput}. This yields the spatial distributions of  the different emission components, as well as the column density in each pixel. Figure \ref{fig:templatesmap} shows the   maps obtained for \object{NGC~7023~NW}. Maps for the other PDRs are displayed in Fig.\,\ref{fig:allmaps}. The spatial distributions of the PAH$^+$, PAH$^0$ and eVSG components are  found to be close to those reported in previous studies \citep{rapacioli05, berne07a, rosenberg11}. 
 The comparison between the distributions of $A_{\rm V}$ and the AIB emission is consistent with the interpretation that the mid-IR emission is dominated by the excited edge of a dense structure (cloud or filament). 
The  column densities determined with this method are  in line with those derived from other tracers \citep[see for example][]{habart03, gerin98} and agree in absolute values within a factor of 4. 
The spatial patterns of the $A_V$ maps seem indeed to be associated with variations  in the physical or geometrical properties. For instance, the peak of column density observed in \object{NGC~7023~NW}  perfectly matches the high-density region traced by CS emission (Pilleri et al., in prep.).

\subsection{PDRs with mixed mid-IR populations}

 Figure \ref{fig:spectra_aib} shows the observed spectra of the PDRs with mixed mid-IR populations and their fits, whose main results are reported 
in Table \ref{table_unresolvedresults}. The quality of the fit can be evaluated using the signal-to-noise ratio (S/R), 
defined as the ratio between the power of the fit divided by the power of the residuals (see Table \ref{table_unresolvedresults}). 
All S/Rs are above 100 (20 dB), which attests to the good quality of the fit. The extinction we derive is compatible with previous estimates that used molecular 
tracers or (sub-)millimeter dust emission \citep[see, for example, ][]{tauber94, kemper99, habart05, karr05}. The weights of the different populations vary 
in the PDRs, reflecting a difference in chemical abundances. PAH$^x$ were suggested to be present only in highly excited regions 
\citep{joblin08, berne09}, and not in mild UV-field reflection nebulae \citep{rapacioli05, berne07a}. Indeed, PAH$^x$ represent a minor fraction of the AIB emission ($\lesssim$10\%) except for the nucleus of the starburst galaxy M82.
\begin{table*}[thdp]
{\small
\begin{center}
\caption{Results of PAHTAT for the PDR with mixed mid-IR populations.}
\label{table_unresolvedresults}
\begin{tabular}{c|cccc|cccc|ccc}
\toprule
Source & 			\multicolumn{4}{c}{Integrated Intensities Ê[W\,m$^{-2}$\,sr$^{-1}$]}	& \multicolumn{4}{c}{PAH and eVSG fractions Ê[\%] Ê \tablefootmark {(a)}}& \multicolumn{3}{c}{Results}		\\
\midrule
 Ê Ê 	& 	 Ê $I_{\rm obs}$		& 	$I_{\rm bands}$ & Ê$I_{\rm continuum}$	& 	$I_{\rm gas}$	& PAH$^0$ Ê& ÊPAH$^+$ 	& PAH$^x$ Ê Ê 		 Ê & Ê eVSG	 Ê Ê & S/R\tablefootmark{(b)} Ê& $f_{\rm eVSG}$\tablefootmark{(c)} 	& $A_{\rm V}	 $	\\
\midrule
Horsehead	& 	$8.3\times10^{-6}$	& Ê$4.1\times10^{-6}$	& 	$3.7\times10^{-6}$	& 	$3.9\times10^{-7}$		& 	32	& 	2	& 	1	& 64	 Ê&238 &0.78	& 6.5	\\
Ced 201		& 	$2.4\times10^{-5}$	& Ê$1.6\times10^{-5}$	& 	$7.6\times10^{-6}$	& 	$1.2\times10^{-7}$		& 	21	& 	21	& 	11	& 45	 Ê&281 & 0.63	& $\lesssim 3$	\\
IC 63		& 	$7.2\times10^{-6}$	& $4.2\times10^{-6}$	& 	$2.8\times10^{-6}$	& 	$2.1\times10^{-7}$		& 	54	& 	0	& 	0	& 45	 Ê& 111& 0.63	& $\lesssim 3$\\
P18 S		& 	$4.8\times10^{-5}$	& $3.2\times10^{-5}$	& Ê Ê	$1.5\times10^{-6}$	& 	$3.5\times10^{-7}$		& 	48	& 	18	& 	10	& 22	 Ê&303 & 0.37	& 10.0\\
P18 N		& 	$5.1\times10^{-5}$	& $3.4\times10^{-5}$	& Ê Ê	$1.5\times10^{-5}$	& 	$5.0\times10^{-7}$		& 	57	& 	13	& 	11	& 17	 Ê&173 & 0.30		& 21.2\\
Orion Bar		& 	$9.9\times10^{-3}$	& $3.9\times10^{-3}$	& Ê Ê	$5.8\times10^{-3}$	& 	$4.8\times10^{-5}$		& 	59	& 	32	& 	0	& 8 Ê &161 & 0.15		& 8.9\\
\bottomrule
\end{tabular}
\tablefoot{
\tablefoottext{a}{Relative to $I_{\rm bands}$.}
\tablefoottext{b}{$S/R = \int{I_{model}^2(\lambda)}/\int({I_{model}(\lambda)-I_{obs}(\lambda))^2}$}
\tablefoottext{c}{Fraction of C atoms locked in eVSGs. See Eq. \ref{fevsg} for details.}
}
\end{center}
}
\end{table*}

\section{Linking the chemistry with the physical conditions}
\label{linking}

 \citet{rapacioli05} and \citet{berne07a}  have suggested
that  eVSGs might be PAH clusters that evaporate into free-flying PAH molecules when  exposed to UV radiation in PDRs.
 In the following, we assume that all  carbon contained in  eVSGs is transferred into PAHs, i.e. that the number of C atoms per unit volume 
contained in AIB carriers, $n_{\rm C}=n_{\rm C}^{\rm eVSG} + n_{\rm C}^{\rm PAH}$, is conserved.
 We have shown in \citet{joblin10}  that there are two independent ways to derive the abundance of C in the mid-IR carriers, either by analysing the intensity of the mid-IR emission or by using the total column density derived from the fit. We have checked that the ratio of these two quantities varies by less than 10\% in the cut of NGC7023 NW shown in Fig.\,\ref{fig:pdrprofiles}, excluding strong variation in the abundance of carbon at the transition eVSG/PAH.

In  Appendix \ref{sec:enbalance} we show that the emissivity per carbon atom of eVSGs and PAHs can be considered as similar. 
This means that $f_{\rm eVSG}$  directly 
represents  the fraction of carbon atoms locked up in eVSGs compared to all carbon atoms in the AIB carriers.  Therefore, we can quantify the evolution of the abundance of eVSGs as a function of the physical conditions ($G_0$ and the gas density  $n_{\rm H}$) by using the quantity $f_{\rm eVSG}$, which can be directly derived from the observations (cf. Equation \ref{fevsg}). 

\begin{figure*}[t]
\centering
\includegraphics[angle = -90, width = .7\hsize, trim=0cm 0cm 0cm 0cm]{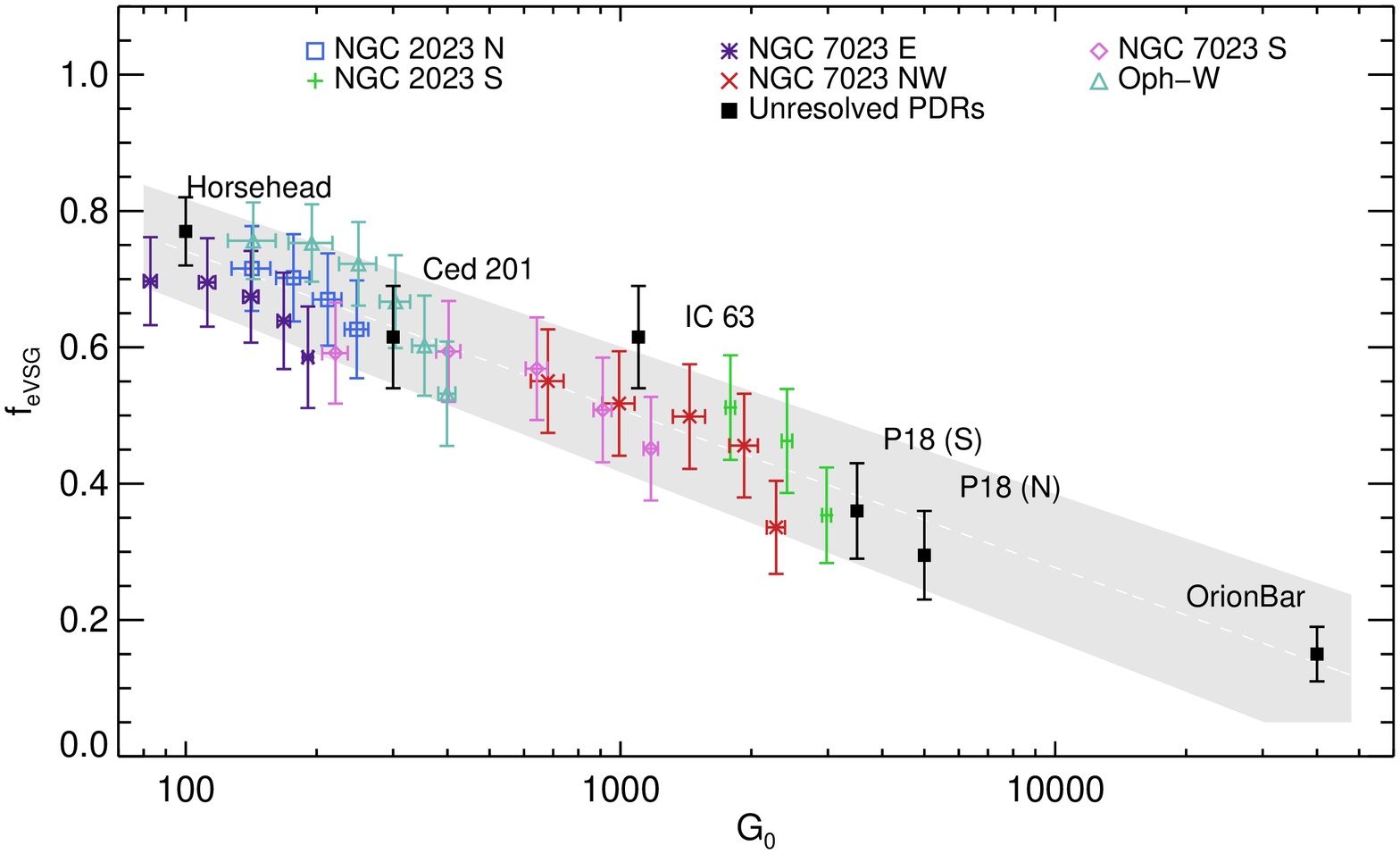}
\caption{Fraction of carbon atoms locked in eVSGs relative to the total carbon in the AIB carriers (derived with PAHTAT) as a function of the intensity of the local UV radiation field. 
 The error bars result from the propagation of the uncertainty on the intensity of the eVSG continuum (see Sect. \ref{section_contcorr}).}
\label{fig:fcprofiles}
\end{figure*}

\subsection{The relation between $G_0$ and $f_{\rm eVSG}$}

 For each of the spatially resolved PDRs, we focused on one cut that extends perpendicularly from the star's position to the PDR and a few arcsecs past the PDR front (see Figs.~\ref{fig:templatesmap} 
and \ref{fig:allmaps}). 
Qualitatively, the variation of $f_{\rm eVSG}$, calculated using Eq.\,\ref{fevsg}, along each of the cuts shows that the eVSG abundance increases from the PDR front into the inner layers and reaches its maximum just behind the  front. The carbon content in eVSGs decreases significantly in the more exposed part of the PDRs, confirming the photo-evaporation scenario. 

To quantify this process, we need an estimate of the value of $G_0$ at various depths of the PDR along this cut, which is not known {\it a priori}. 
This was performed using the approach presented in \citet{habart03} for the Oph-W filament. This method relies on the modelling of the AIB emission using a few physical and geometrical assumptions\footnote{For the sake of completeness and to facilitate reading, the description of this method is reported in Appendix \ref{sec:energetics}.}. 
This  yields the values of $G_0$ and $n_{\rm H}$ at various depths in the PDR (upper panels in Fig. \ref{fig:pdrprofiles}), which can be compared with the variation of $f_{\rm eVSG}$ (lowest panels in Fig. \ref{fig:pdrprofiles}).

In Fig. \ref{fig:fcprofiles}, we report the values of $f_{\rm eVSG}$ as a function of the intensity of the UV  radiation field $G_0$ for all  regions. For each cut, the number of points that are reported depends on the spatial resolution of the observations.  In each spatially resolved PDR, we excluded the low-density ($n_{\rm H}\lesssim5\times10^{3}$\,cm$^{-3}$) layers that lie in the most exposed region of the PDR (up to an $A_{\rm V} \sim 0.1-0.3$).  In these regions, eVSGs are underabundant because the photoevaporation process is expected to be much faster than reformation. Hence their contribution to the mid-IR emission is minor and the derivation of their abundance is subject to large uncertainties. Instead, we concentrated on the PAH-eVSG transition, which arises at  $A_{\rm V} \sim 0.5$ and  is spatially coincident with the H$_2$ pure rotational emission.
 This corresponds to regions in which fresh matter from dense regions is exposed to UV photons. Finally, we also added  one point for each of the PDR
with  mixed mid-IR populations, for which we relied on the values of $G_0$ and $n_{\rm H}$ that are  commonly used in the literature (Table \ref{tab:otherpdrs}). 

 Figure \ref{fig:fcprofiles} shows  a clear decrease of the fraction of carbon in eVSGs with increasing intensity of the UV radiation field. The trend can be fitted by the expression

\begin{equation}
 f_{\rm eVSG} = (-0.23 \pm 0.02) \log_{10}(G_0)+(1.21\pm0.05).
\label{eq:fc}
 \end{equation}
 
 This implies that $f_{\rm eVSG}$ can be used to trace the  (local) intensity of the UV radiation field.  At low UV field intensities ($G_0 \lesssim 100$),  the mid-IR fits are less reliable because the AIB emission is fainter and sometimes barely detected, therefore the $f_{\rm eVSG}$  vs  $G_0$ relationship cannot be established. Similarly, the extrapolation of the relation between 
$f_{\rm eVSG}$ and $G_0$ to more intense UV fields (e.g. above $5\times 10^4$, which is our limit in this paper) is also difficult because in these environments, the photo-processing 
of the eVSG and PAH populations is extreme and leads to the difficulty to observe the more fragile eVSG species, and likely breaks the assumption on the conservation of 
$n_{\rm C}$ in the AIB carriers. We conclude that the procedure that is described in this paper to derive $G_0$ is reliable in the range  $100 \lesssim G_0 \lesssim 5\times 10^4$.
The gas densities that have been explored span the range $10^2$ - $10^5$\,cm$^{-3}$.

\subsection{Correlation with other tracers}

\begin{figure}[t]	  
\centering
\includegraphics[angle =270, width =0.98\hsize, trim = 0cm 0cm 0cm 0cm]{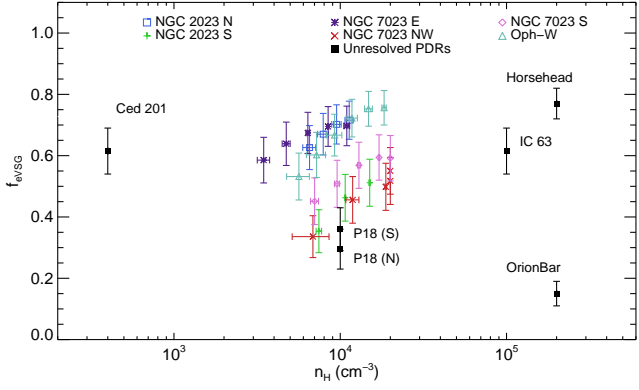}
\caption{Relation between the local density and the $f_{\rm eVSG}$ value for the PDRs with spatially
resolved ( Table~\ref{tab:pdrinput}) and mixed (Table~\ref{tab:otherpdrs}) mid-IR populations. 
 Error bars are calculated as in Fig.\,\ref{fig:fcprofiles}.}
\label{fig:nhfc}
\end{figure}

Figure \ref{fig:nhfc} displays the correlation between $f_{\rm eVSG}$ and the local density for the PDRs
with both mixed and spatially resolved mid-IR populations, showing that these two quantities are uncorrelated. Similarly, no correlation appears between 
$f_{\rm eVSG}$ and the ratio $G_0/n_{\rm H}$, despite the link of this ratio with the fraction of ionised PAHs by \citet{fleming10} in their detailed study of \object{NGC~7023~NW}, \object{NGC~2023~N}, and \object{NGC~2023~S}. This proves that the 
evaporation of the eVSG population is indeed directly linked to the intensity of the UV radiation field through a
mechanism that differs significantly from those driving the ionizasion balance. This strongly suggests that  the reformation of 
eVSGs is not efficient in these environments.
\citet{rapacioli06} have indeed shown that the photo-evaporation of PAH clusters in   the  NGC~7023~NW PDR is much faster than their reformation by collisions with PAHs.

\subsection{Use of PAHTAT to derive $G_0$ or other parameters}

 In principle, PAHTAT can be used to fit any mid-IR spectrum that shows AIB emission to provide $f_{\rm eVSG}$ and therefore the local value of $G_0$. However, when using this tool, one must bear in mind for which frame  this relationship is applicable. PAHTAT was calibrated in relatively dense and highly UV-irradiated environments that are found in star-forming regions. This implies 1) that the exciting and evaporating radiation field results from massive stars so that
the radiation field is dominated by UV photons, and 2) the presence of dense regions where eVSGs are abundant and start to evaporate where the UV irradiation penetrates. In other types of regions,
 the value of $G_0$ derived from PAHTAT may not be reliable. Still, PAHTAT can be used to identify unusual spectral properties of AIBs as related to peculiar properties of the source. \citet{vega10}  for instance used the PAHTAT templates and  identified an exceptionally intense 11.3 $\mu$m features in early-type galaxies. They have interpreted this as a population of large neutral PAHs, which are not included in our templates.
 
The mid-IR emission of star-forming galaxies mainly arises from unresolved dense and highly UV-irradiated PDRs in which PAHTAT could be used to derive an average radiation field. To test the reliability of the value of $G_0$ derived with PAHTAT on unresolved PDRs, we used the fitting tool to determine a value of $f_{\rm eVSG}$ for the average spectra calculated over more extended fields of view (up to 1 arcmin) in our template object, \object{NGC~7023~NW}. This can be compared with $<f_{\rm eVSG}>$,  the mean value obtained by averaging results on the individual pixels contained in the same region. 
The  two values agree well for regions close to the eVSG peak. For regions close to the PDR front, the value of $<f_{\rm eVSG}>$ is higher but by less than 0.05. This yields an overestimate of the mean UV radiation field by less than 50\,\%. Although this error is not directly applicable to other PDRs, it shows that the method can be used to derive an effective UV radiation field intensity in unresolved sources within a factor $\sim2$.

\begin{figure}[th]
\includegraphics[width = .99\hsize, trim=1cm 2cm 0cm 1cm]{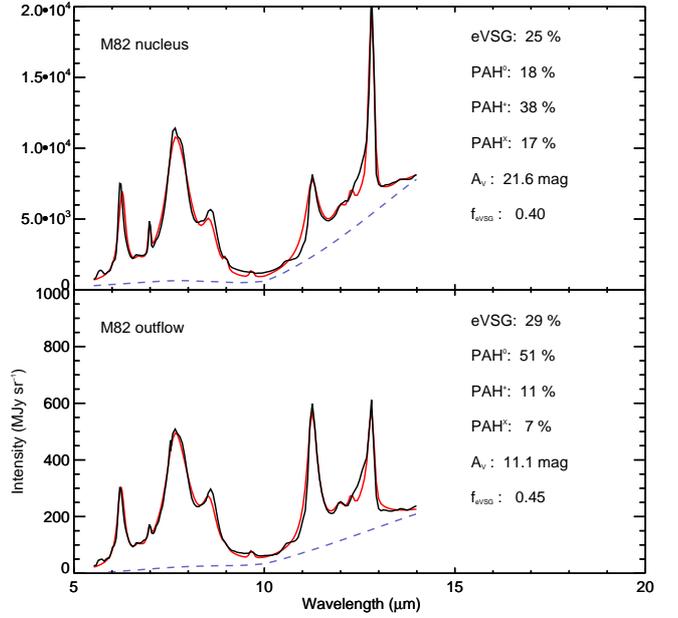}
\caption{Mid-IR spectra (black line) of the nucleus and the outflow of \object{M82} and their fits (red line) obtained with PAHTAT. The results of the fit suggest that the UV radiation field in the nucleus is slightly higher, as well as the column density of the gas.  }
\label{fig:m82}
\end{figure}

As an example, we applied our fitting procedure to the IRS spectra of the  nucleus and  outflow of the starburst galaxy \object{M82}. The fit results are shown in Fig. \ref{fig:m82}. 
The \object{M82} IRS spectral map has been studied previously by \citet{beirao08}. They found that the magnitudes of visual extinction were ranging between $A_{\rm V} = 0$ to 50 and were likely responsible for most of the spectral variations in the AIB features. The $A_{\rm V}$  values derived from our decomposition on each pixel are consistent with these values. The results of the fit applied to the mean spectra of the nucleus and the outflow of M82 are still consistent with the results from \citet{beirao08}, because we derive an estimate of the extinction along the line of sight  of $A_{\rm V}\sim21$ for the  nucleus and $\sim 11$ for the  outflow.   The values of $f_{\rm eVSG}$ derived from the mid-IR fit correspond to a UV radiation field of $G_0 = 3^{+2}_{-1}\times 10^3$  for the nucleus and  $G_0 = (2\pm1) \times 10^3$ for the outflow.  Previous studies using  chemical modelling of the nucleus of \object{M82} by \citet{fuente08} have predicted a value of $G_0 \sim 1 \times 10^4$, whereas those using far-IR observations \citep{colbert99, kaufman99} report lower values ($10^{2.8}\lesssim G_0 \lesssim10^{3.5}$). These agree well  with our estimates.

\section{Conclusion}	 
\label{sec:conclusion}

We have presented a new method for analysing the mid-IR spectra of PDRs using a basis of PAH and eVSG template spectra.  We have shown that with a few, physically meaningful  parameters  it is possible to obtain a very good fit of the mid-IR spectra of PDRs that span a wide range of physical conditions. The fit returns the relative contribution of the PAH and eVSG populations to the mid-IR emission, the intensity of the gas lines (especially the rotational lines of H$_2$), and the magnitudes of visual dust extinction along the line of sight.

We derived the abundance of C atoms locked in eVSGs relative to the total abundance of carbon in the AIB carriers. This value strongly correlates with the local UV field intensity and can therefore be used to quantify this important physical parameter.
Our study has provided new insights into the properties of eVSGs, supporting a chemical scenario in which these species are destroyed by UV photons to produce free-flying PAHs. PAH clusters and PAH-Fe complexes have been proposed as plausible carriers for these eVSGs and  detailed studies on the survival of these species in PDRs have been performed \citep{rapacioli06, simon09}. One important conclusion of our observational study is that eVSGs are destroyed in PDRs faster than they can be reformed. This suggests that they are formed in the denser parts of molecular clouds, although this scenario  will be difficult to probe since in these regions  mid-IR emission cannot be triggered and therefore eVSG fingerprints cannot be observed. Another intriguing aspect is the continuum emission that seems to be associated to eVSGs. We showed that the intensity of this continuum is quite variable relative to the band intensity, likely reflecting  a chemical or  size distribution effect. Solving this question opens new perspectives for studies on laboratory analogues of eVSGs. 

Finally, our  fitting code  PAHTAT (PAH Toulouse Astronomical Templates) is  made publicly available to the community as an IDL package, together with the template spectra used in this work. This procedure combined with the high spatial resolution and sensitivity of forthcoming missions such as JWST and SPICA will allow us to determine the spatial variation of the UV radiation field in more distant galaxies and protoplanetary disks.

\begin{figure*}[ht]	
\centering
\includegraphics[width =0.65\hsize]{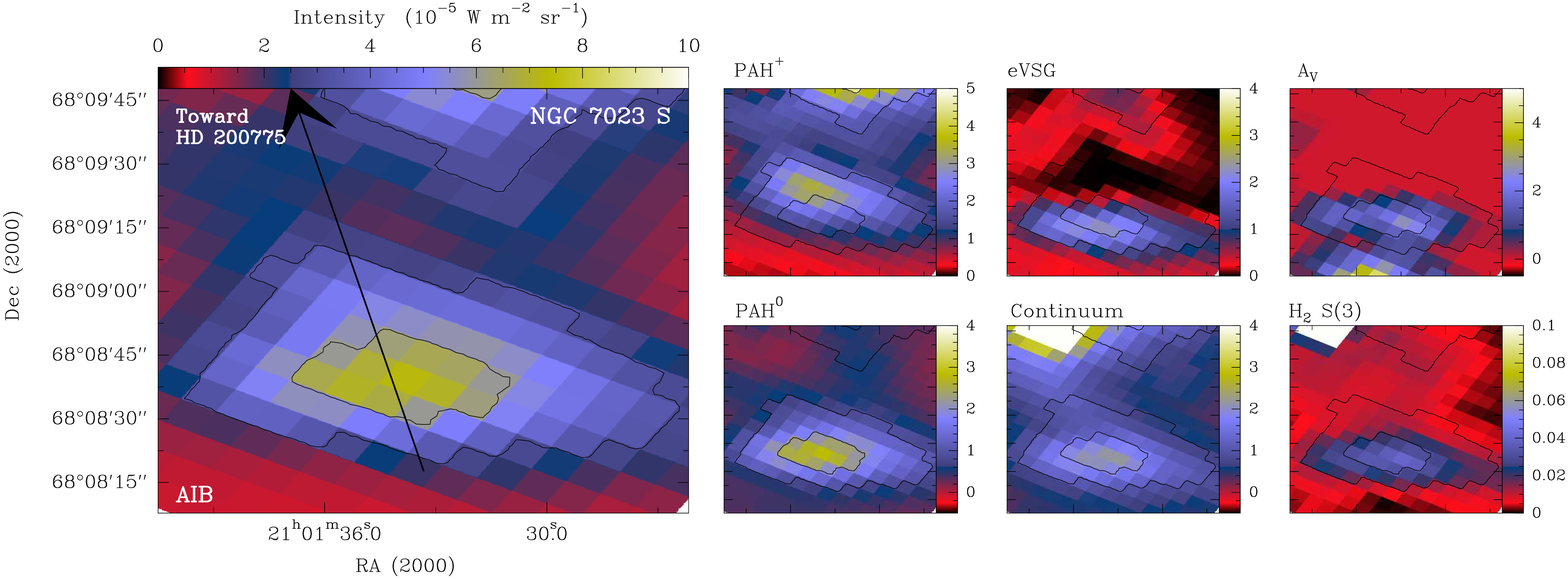}\vspace{0.3cm}
\includegraphics[width =0.65\hsize]{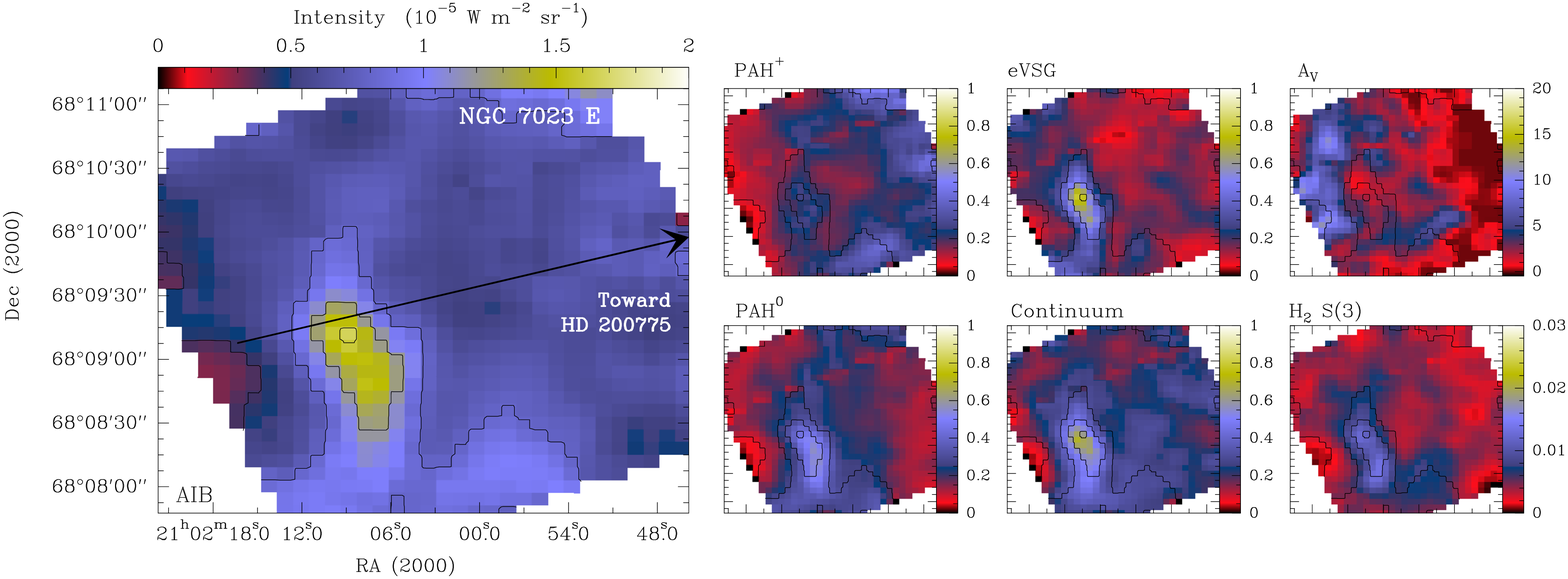}\vspace{0.3cm}
\includegraphics[width =0.65\hsize]{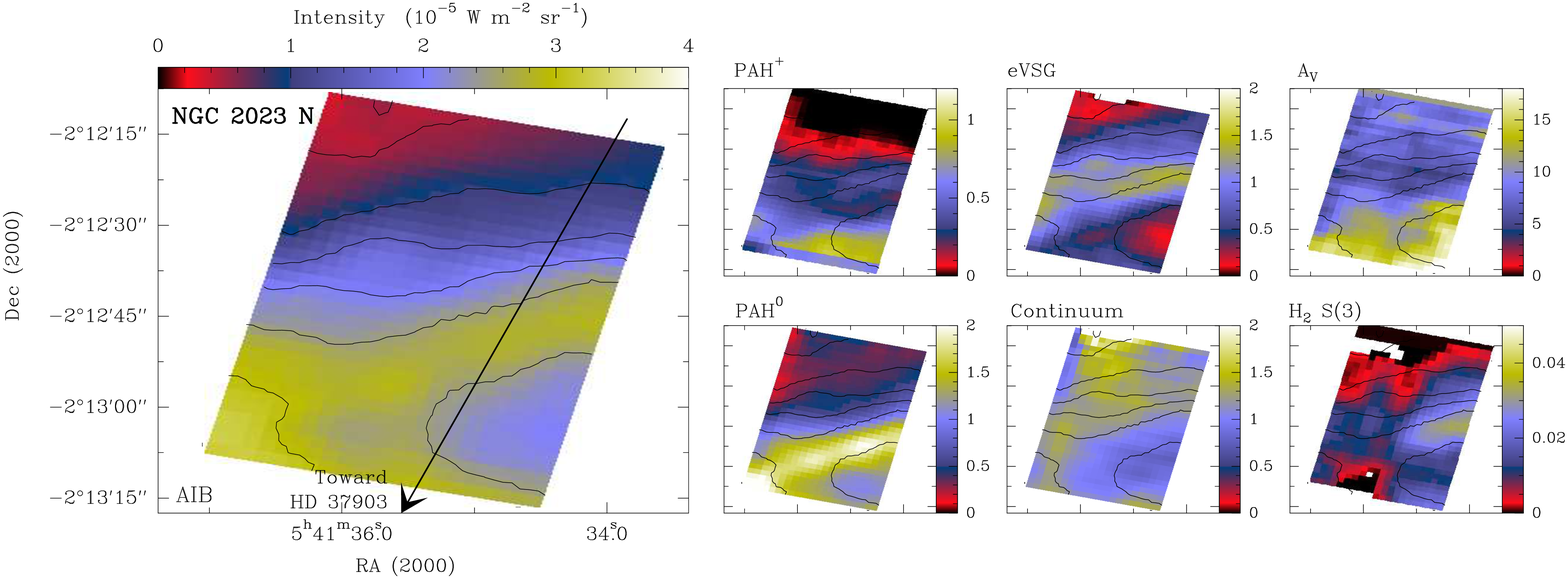}\vspace{0.3cm}
\includegraphics[width =0.65\hsize]{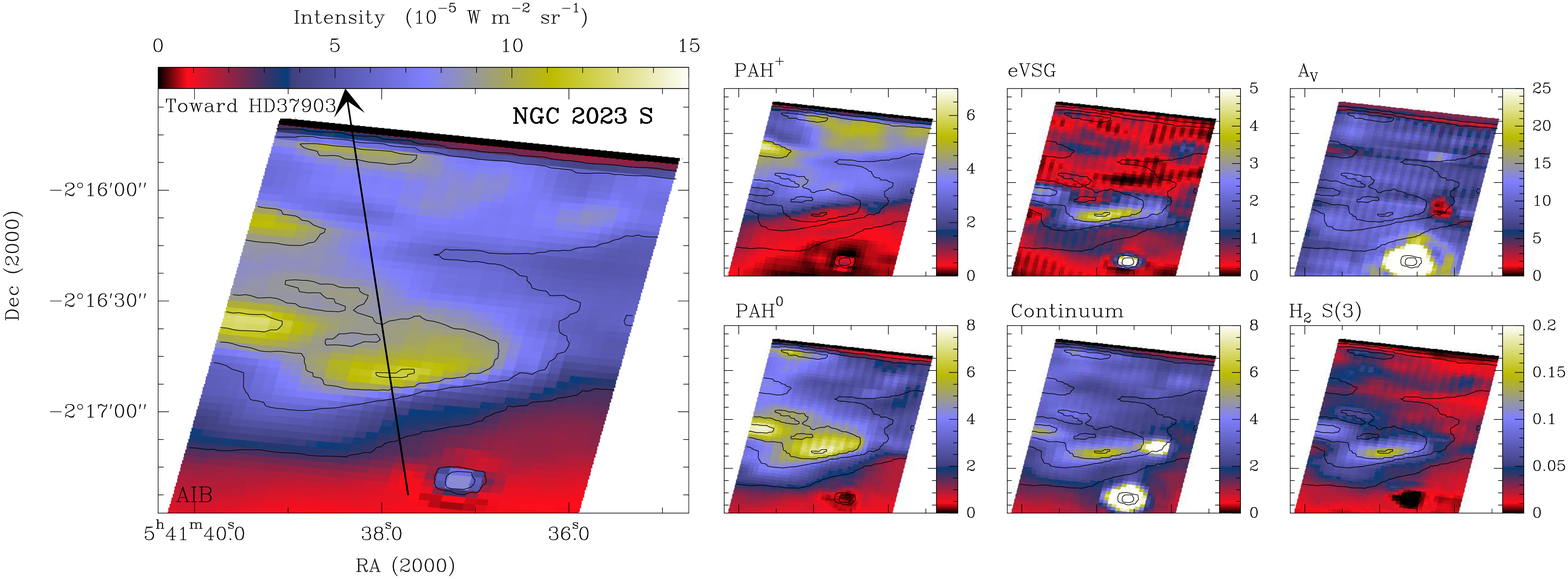}\vspace{0.3cm}
\includegraphics[width =0.65\hsize]{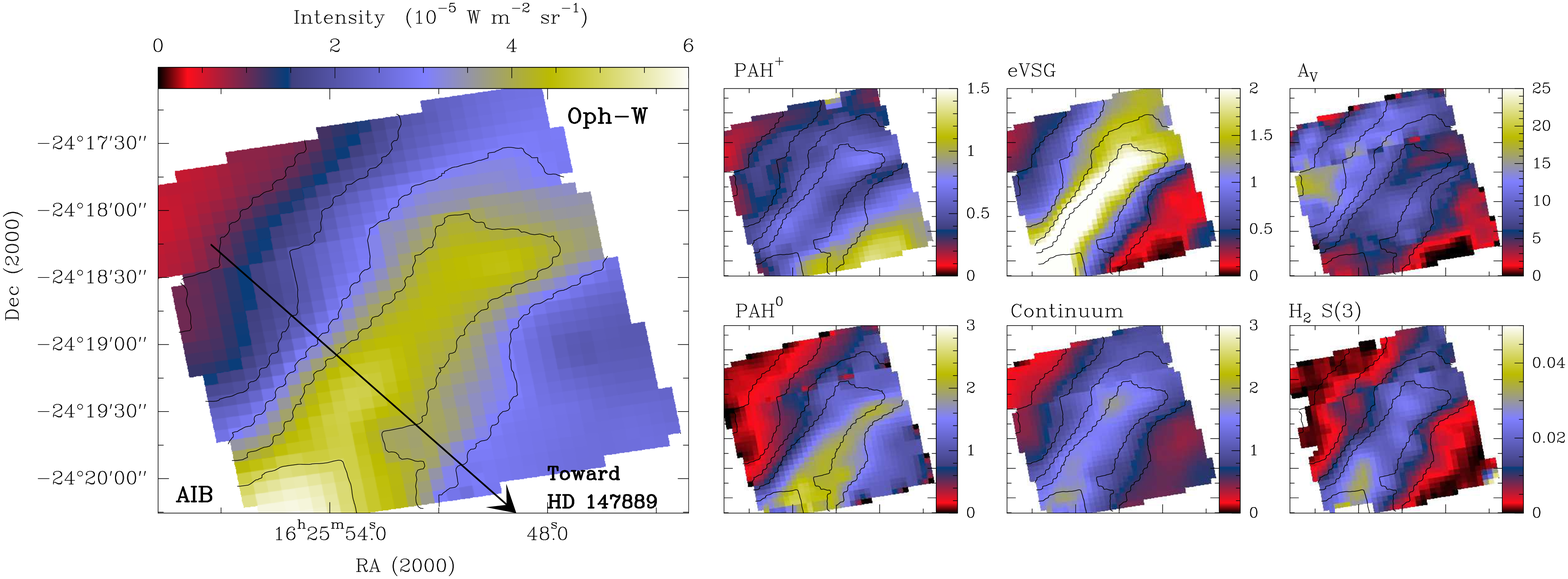}
\caption{Same as figure \ref{fig:templatesmap} but for the other sources considered in this work: from top to bottom, \object{NGC~7023~S}, \object{NGC~7023~E}, \object{NGC~2023~N}, \object{NGC~2023~S},  and the $\rho$-Ophiucus Filament. The solid lines represent the cuts studied in this work, whose analysis is shown in Fig. \ref{fig:pdrprofiles}. }
\label{fig:allmaps}
\end{figure*}

\begin{figure*}[ht]	
\centering
\includegraphics[ width=0.33\hsize]{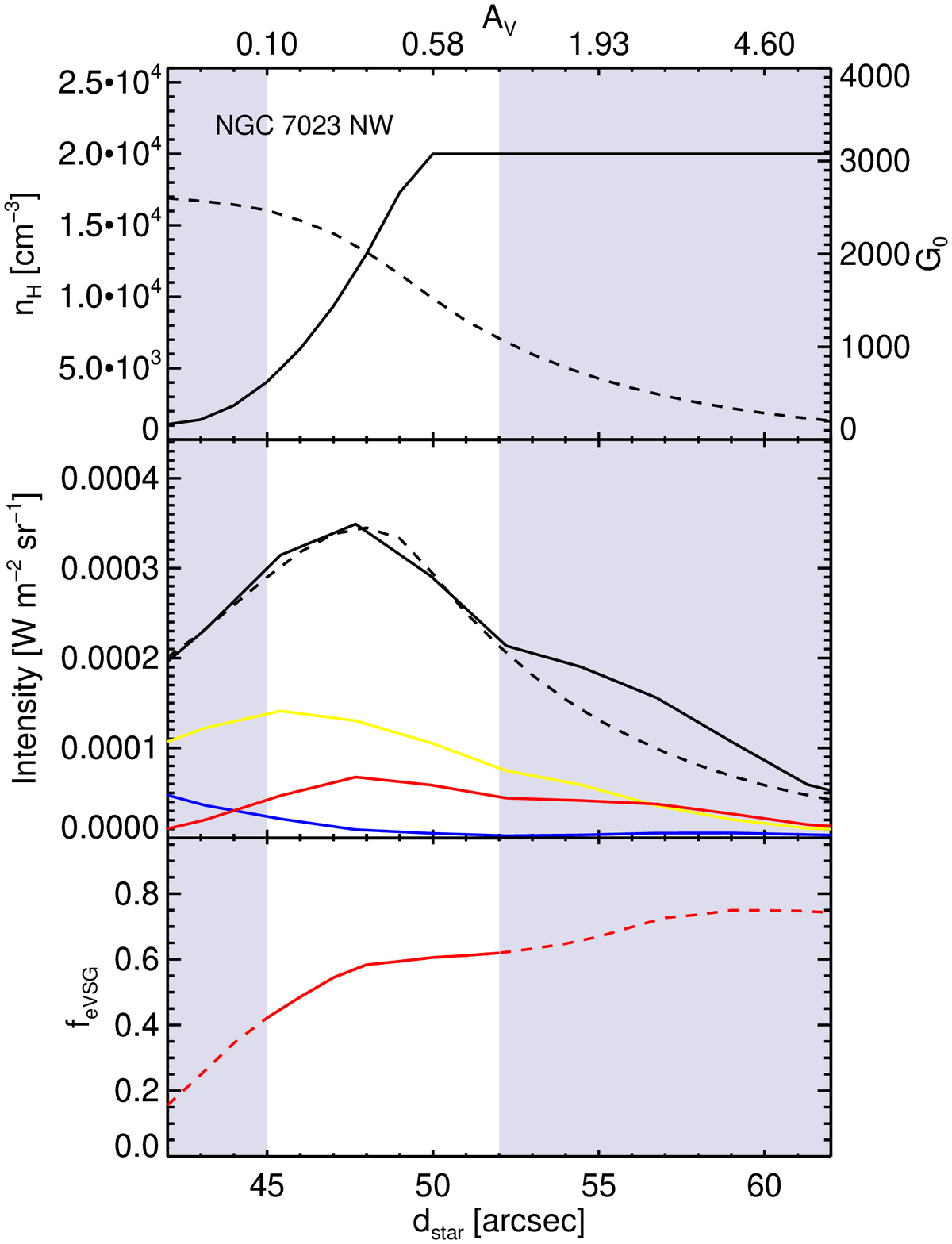}
\includegraphics[ width=0.33\hsize]{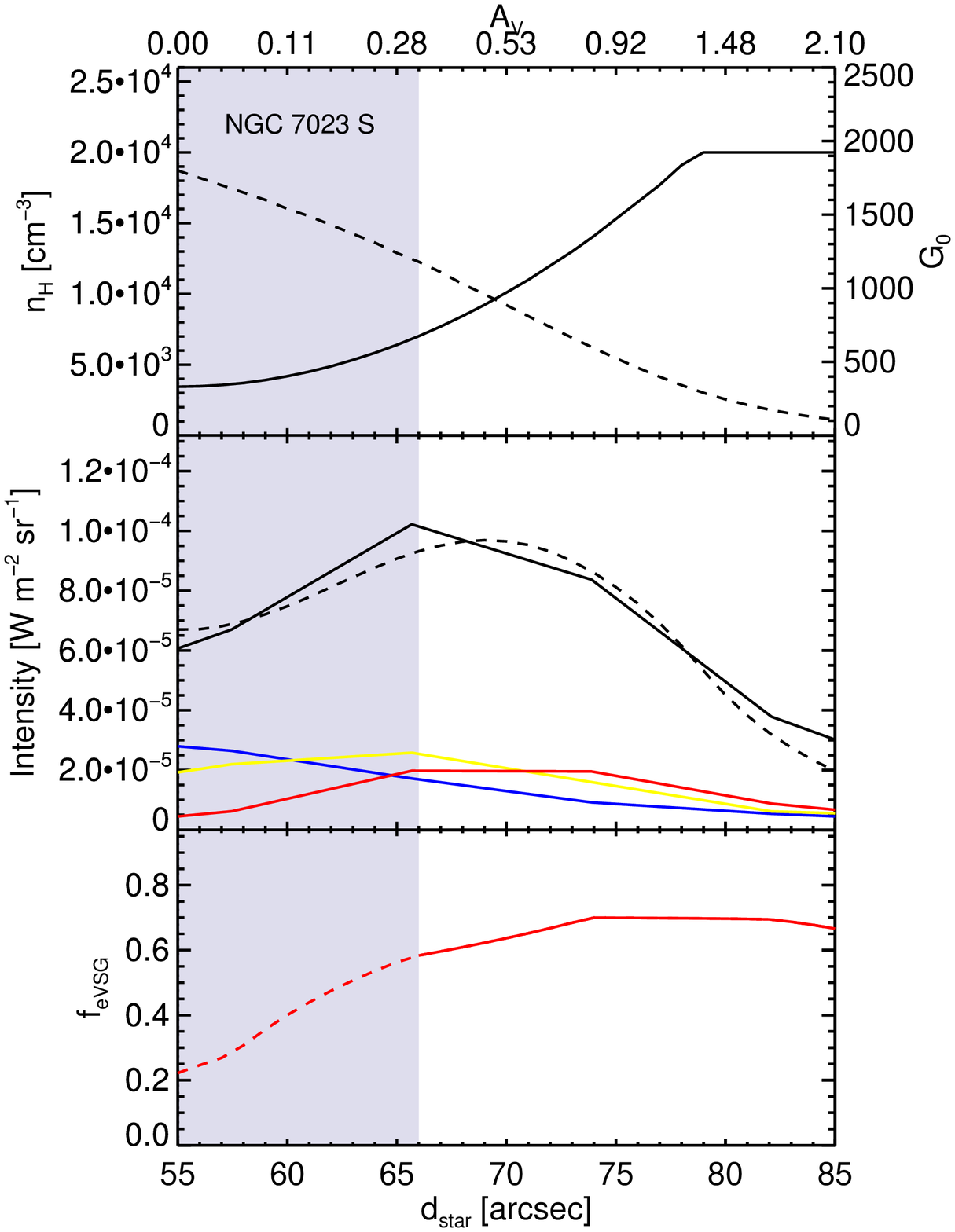}
\includegraphics[ width=0.33\hsize]{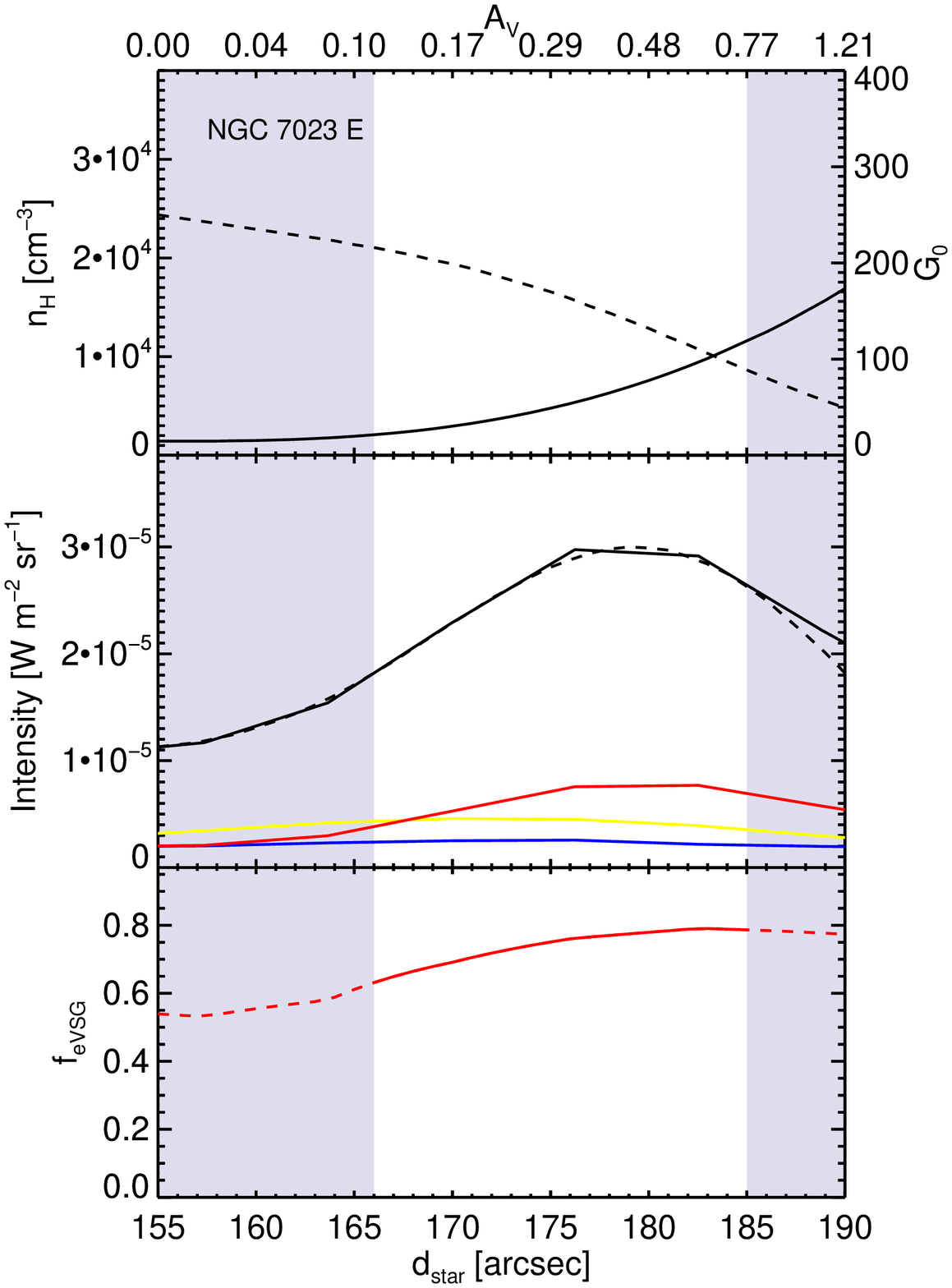}
\includegraphics[ width=0.33\hsize]{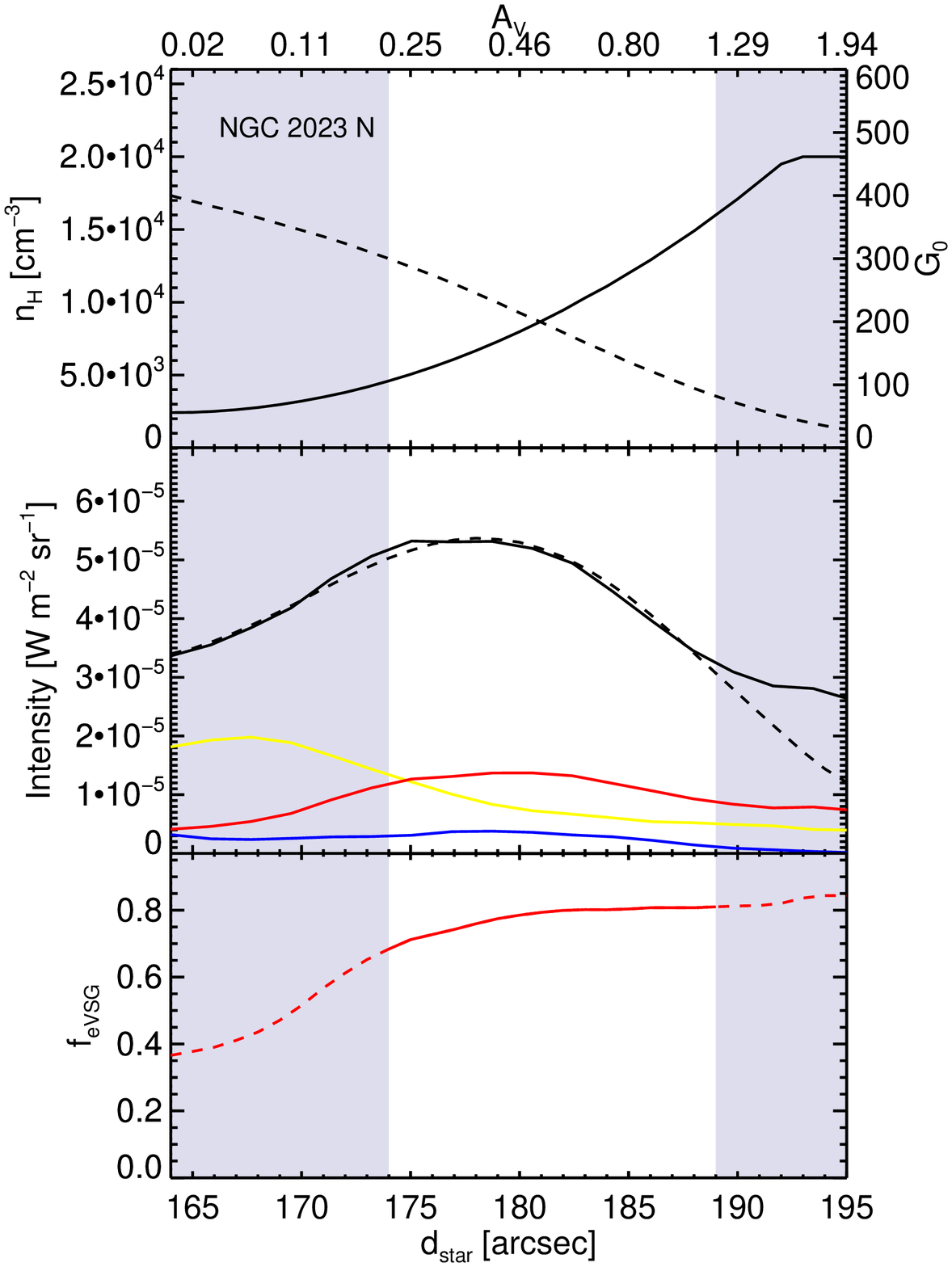}
\includegraphics[ width=0.33\hsize]{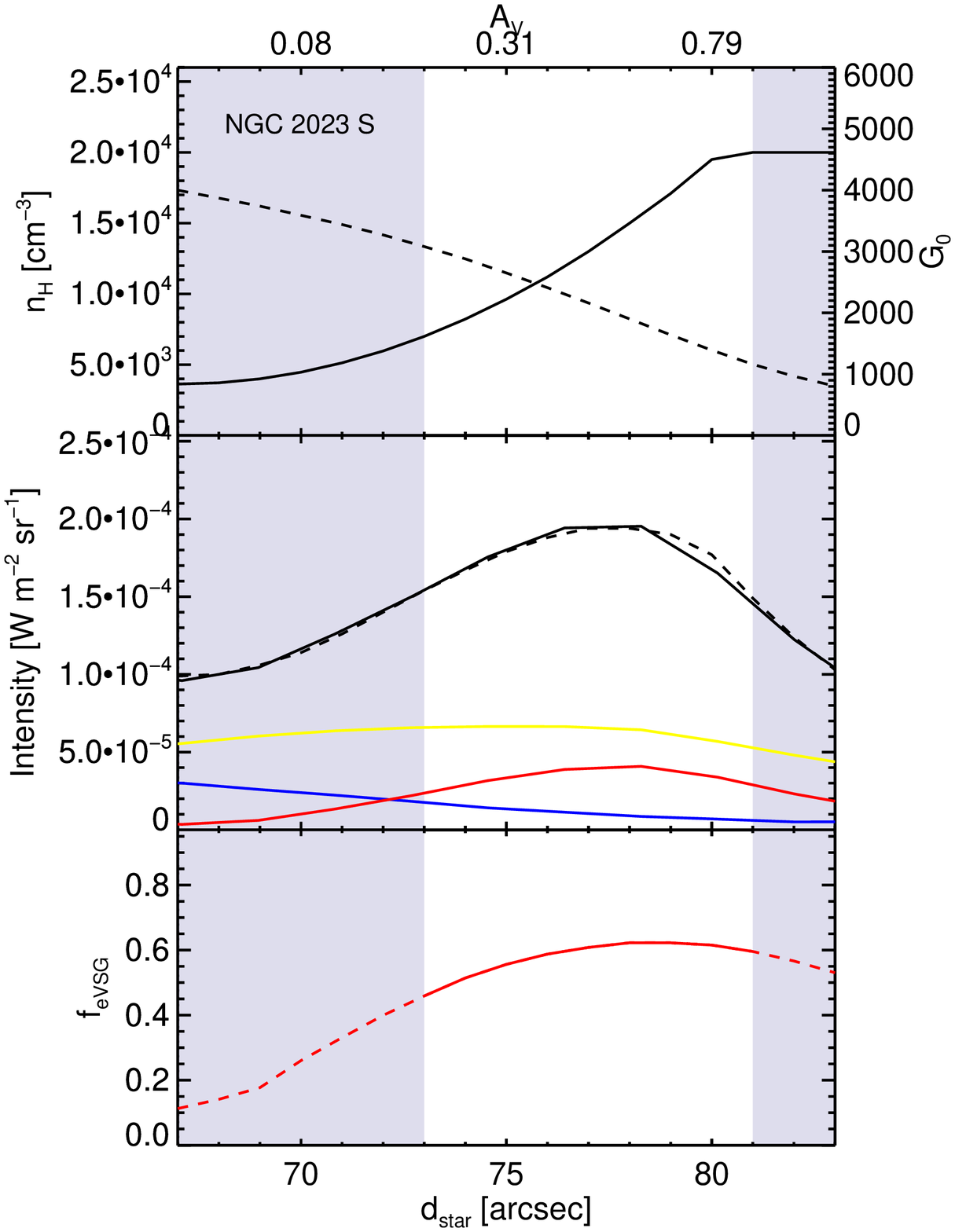}
\includegraphics[ width=0.33\hsize]{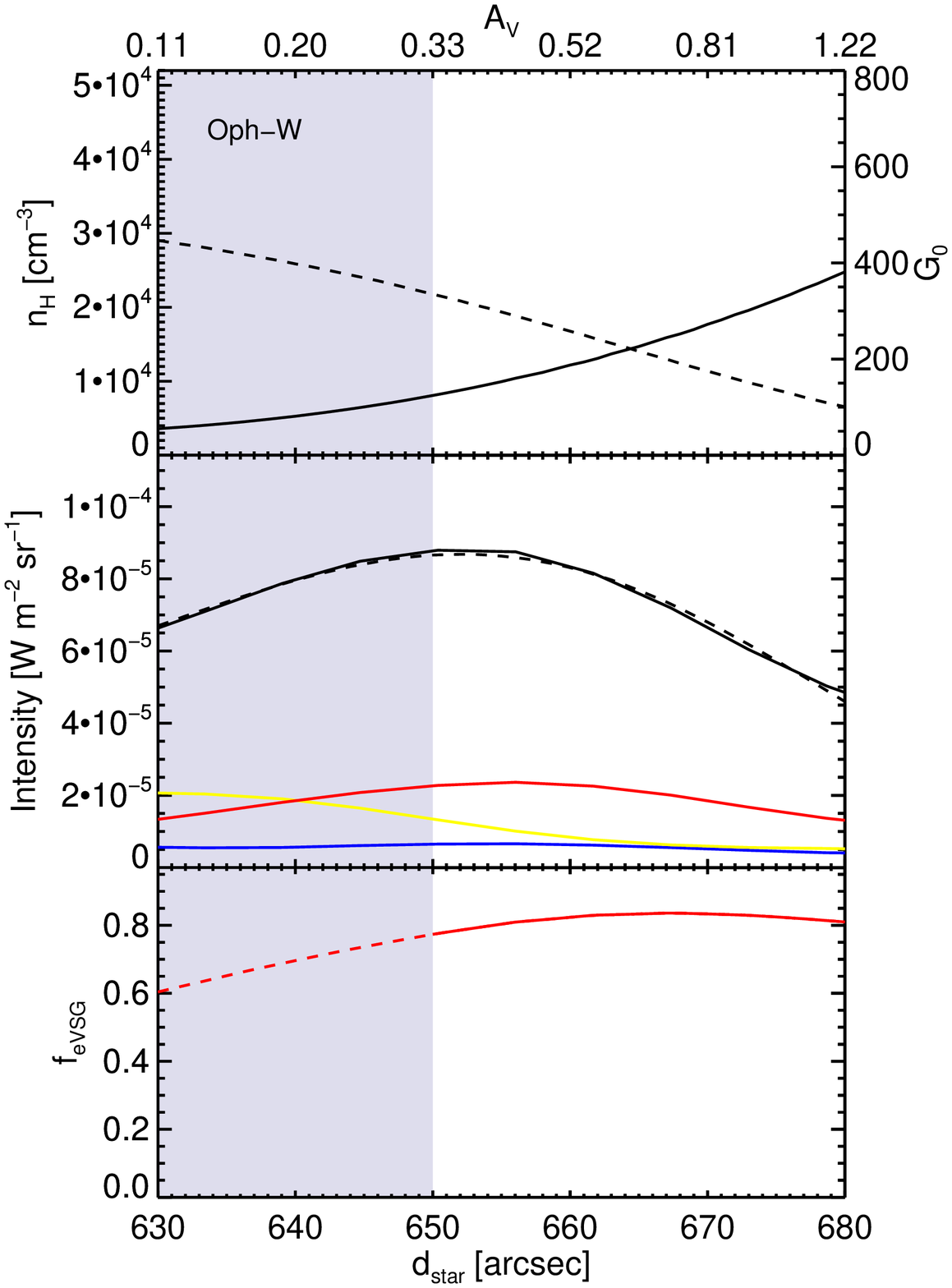}
\caption{Compilation of our results along each PDR cut defined in Figs. \ref{fig:templatesmap} and \ref{fig:allmaps}. {\it Upper panels:}  the local density and UV radiation field
profiles as a function of the distance from the star along the $r$ or
$A_{\rm V}$ axis of Fig. \ref{schema_geo}.  {\it Central panels}:  Blue, yellow, and red represent the  contribution of PAH$^+$, PAH$^0$ and eVSGs respectively, as extracted from the  fit. Solid black lines represent the corrected mid-IR emission profile. The dotted lines represent the fit obtained with the geometrical model described in Appendix \ref{sec:energetics}. {\it Lower panels: } the variation in the fraction of carbon atoms locked in eVSGs relative to the total carbon in  all the AIB carriers.  The grey-shaded regions are those that   were not  considered in  Fig.\,\ref{fig:fcprofiles}. and Eq.\,\ref{eq:fc}.}
\label{fig:pdrprofiles}
\end{figure*}

\begin{appendix}

\section{Mid-IR emission model} \label{sec:energetics}

In this section, we describe how we derived the local physical parameters, i.e. the gas density ($n_{\rm H}$) and the UV radiation field ($G_0$) from the local mid-IR emission of the AIB carriers.

\subsection{Energy balance}
\label{sec:enbalance}

The UV absorption cross-section of PAHs was measured to be proportional to the number of involved carbons  \citep{joblin92}. This has also been shown on a larger sample of species using theoretical calculations\footnote{http://astrochemistry.ca.astro.it/database/} \citep{malloci04, malloci07}. 
Using a typical value for $\sigma_{C}^{\rm PAH}$, one can estimate that screening atoms inside a grain by surface atoms is negligible for grains with lesser than $\sim 5 \times 10^4$ carbon atoms.
As mentioned in Sect. \ref{linking}, eVSGs considered in these work are relatively small, containing up to about 1200 carbon atoms ($\sim 10$\,\AA), and we therefore considered that the absorption cross sections per C atom of eVSGs and PAHs are similar: 
$\sigma_{C}^{\rm eVSG} = \sigma_{C}^{\rm PAH} = \sigma_{C}$. 

The power absorbed by the AIB carriers  can be expressed as
\begin{equation}
   \label{eq:Eabs}
   P_{\rm abs} = \int_{91.2nm}^{\lambda_{\rm{max}}} n_{\rm C}\sigma_{C}(\lambda) F(\lambda)d\lambda,
\end{equation}
\noindent where $F(\lambda)$ is the flux of the radiation field from the nearby star. Since ionised PAHs and large neutral PAHs (and hence eVSGs) can absorb significantly in the visible \citep[][]{salama96}, 
we assume $\lambda_{\rm{max}} = 1000 $~nm.

As discussed by different authors \citep{boulanger88, habart03}, the power emitted by the AIB carriers can  be expressed as a function of $G_0$:
\begin{equation}
   \label{eq:J}
   P_{\rm em} =  \epsilon \times n_{\rm H}\times G_0,
\end{equation}
\noindent where $G_0$ denotes the UV radiation field in Habing units.
Considering that all species absorb with the same $\sigma_{\rm C}$ and applying  the conservation of energy, an emissivity per C atom of the AIB carriers can be defined as $\epsilon_{C}^{\rm eVSG} = \epsilon_{C}^{\rm PAH} = \epsilon_{C}$. The emissivity $\epsilon$ can be conveniently expressed as 
$$\epsilon=n_{\rm C}\,\epsilon_{\rm C}/n_{\rm H}$$
where $n_{\rm C}$ is the number of C atoms per cubic centimeter that is locked up in the AIB carriers.
Writing the energy balance $P_{\rm em}=P_{\rm abs}$ leads to
\begin{equation}
   \label{eq:eps}
   \epsilon = n_{\rm C}/n_{\rm H} \frac{1}{G_0} \times \int_{91.2nm}^{1000 nm}\sigma_{C}(\lambda)F(\lambda)d\lambda.
   \end{equation}

We computed $\epsilon$ for a variety of PAHs  of different sizes and charges using the values of $\sigma_{\rm C}$ available
from the on-line database of \citet{malloci07}. The results scatter within 10\%. Assuming a value of $n_{\rm C}/{n_{\rm H}}= 4.2\times10^{-5}$
\citep[][for $R_{\rm V}=5.5$]{draine03} we obtain $\epsilon \sim \,5 \times 10^{-32}$ W\,H$^{-1}$, consistent 
with previous estimates by \citet{habart01} and \citet{habart03}. 
Eq.~\ref{eq:eps} shows  that the value of $\epsilon$ 
depends {\it a priori} on the spectral shape (hardness) of the radiation field. To estimate the variation of $\epsilon$ with the 
hardness of the radiation field, we calculated it with the appropriate Kurucz spectra for the stars of Table \ref{tab:pdrinput}. Our calculations show that for the objects studied in this paper, $\epsilon$ marginally depends on the hardness of the radiation field.

\begin{figure}[t]
\centering
\includegraphics[angle = -90, width =\hsize, trim=0cm 2cm 0cm 4cm]{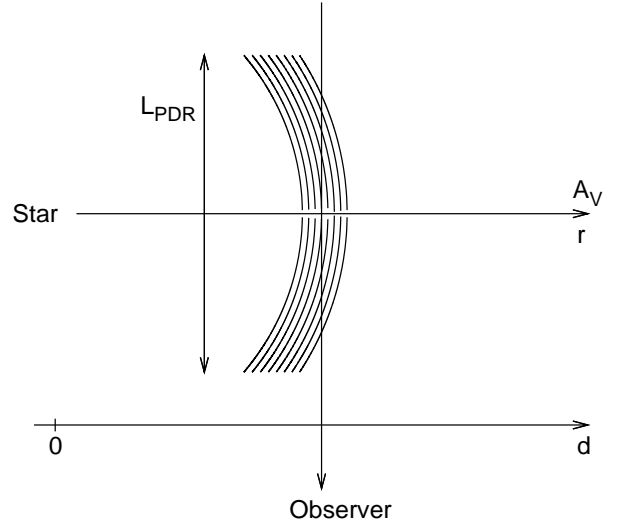}
\caption{Spherical shell geometry used to model the PDRs with spatially resolved mid-IR populations (Table \ref{tab:pdrinput}). The
$r$ and $A_{\rm V}$ axis refers to the local position inside the PDR, along the
star-PDR axis, counted  from the star and from the PDR
front, respectively. The $d$ axis refers to the position of the line of sight on the sky, counted from the star position.}
\label{schema_geo}
\end{figure}

\begin{table*}[thdp]
\begin{center}
\caption{Parameters for the gas density profiles in the PDRs with spatially resolved mid-IR populations, extracted from our emission model as explained in Appendix \ref{subsec:profiles}. References: (a) \citet{gerin98} (b) \citet{fuente95} (c) \citet{habart03}}
\label{tab:fit_results}
\begin{tabular}{c|cc|cccc}
\toprule
Object         &   $d_{\rm front}$   & $n_{\rm H}^{\rm plateau}$ 	& $n_{\rm H}^0$       		   &   $\alpha$ & $d_0$   &  $L$         \\
              	   & [pc]			& [cm$^{-3}$]  				&    [$10^3$ cm$^{-3}$]           &			&  [pc]  	 &     [pc]         \\
\midrule
\object{NGC~7023~NW}    &   0.087     &  $2\times 10^4 {}^{(a)}$   & $1.1$            &   2.0     & 0.016   &   0.13   \\
\object{NGC~7023~S}    &  0.11     &  $2\times 10^4 {}^{(a)}$ & $3.4$            &   2.0     & 0.054   &   0.10          \\
\object{NGC~7023~E}    &  0.32     &  $3\times 10^4 {}^{(a)}$   & $1.4$            &   2.7     & 0.091   &   0.27       \\
\object{NGC~2023~N}      &  0.38     &  $2\times 10^4 {}^{(b)}$ & $2.4$            &   2.0     & 0.069   &   0.27        \\
\object{NGC~2023~S}    &  0.13   &  $2\times 10^4 {}^{(b)}$ & $3.6$            &   2.0     & 0.033   &   0.07          \\
Oph-W          &  0.40     &  $4\times 10^4 {}^{(c)}$   & $2.6$            &   2.5     & 0.051   &   0.28       \\
\bottomrule
\end{tabular}
\end{center}
\end{table*}

\subsection{Determining of the structure of the PDRs with spatially resolved mid-IR populations}
\label{subsec:profiles}

The concepts presented in the previous section can be used to study the physical conditions in PDRs with spatially resolved mid-IR populations (Table \ref{tab:pdrinput}). 
However, to directly relate the observations to the local UV radiation field and gas density, some knowledge on the geometry of the PDR is needed. 
We assumed a spherical shell model as in \citet[][see also figure \ref{schema_geo}]{habart03}, in which the cloud is divided into successive layers of increasing density. 
The star is placed at the centre of the cavity, at a distance $d_{\rm front}$ that is determined by the projected distance on the sky (cf. Table \ref{tab:pdrinput}). 
The total intensity radiated by AIB carriers as a function of the position on the sky is  obtained by integrating Eq.\,\ref{eq:J} along a given line of sight, i.e.:

\begin{eqnarray}	\label{eq:intensity}
 I(z) & = &\frac{1}{4\pi} \int_{0}^{L} \epsilon \times n_{\rm H}(r) \times G_0 (r) \,dl,   \nonumber 
 \end{eqnarray}
where $z$ is the projected distance from the front, $r$ is the distance from the star, and $L$ is the thickness of the PDR perpendicularly to the plane of the sky. 
The local radiation field inside the PDR depends on the impinging radiation field at the PDR surface and on the gas density profile that determines its attenuation. It evolves as
\begin{equation}
G_0(r) = G_0^{\rm front} \times e^{-\sigma_{\rm UV}  \int_0^{r} n_{\rm H}(r) dr},
\end{equation}
where $\sigma_{\rm UV}$ is the dust extinction cross-section at 1000 \AA \,\citep[$1.5\times10^{-21}$ cm$^2$ H$^{-1}$, ][]{weingartner01}. 
The density profile is modelled between the front position at $r=d_{\rm front}$ and a cutoff position $r=d_{\rm front}+d_0$ as a power law from the density $n_{\rm H}^0$ in the diffuse region close to the star, up to a maximum density $n_{\rm H}^{\rm plateau}$. Beyond the cutoff position, the density is assumed to be constant, equal to $n_{\rm H}^{\rm plateau}$:

\begin{eqnarray}	\label{eq:density}
   n_{\rm H} (r) & = & n_{\rm H}^0 + n_{\rm H}^{\rm plateau} \times \left[ \frac{r-d_{\rm front}}{d_0} \right]^\alpha  ~~~ \rm{if~} d_{\rm front} < r < d_0  \\
           & = & n_{\rm H}^{\rm plateau}  ~~~~~~~~~~~~~~~~~~~~~~~~~~~~~~~~~~~~~~~~~~~~~~ \rm{if~} r > d_0,
\end{eqnarray}

\noindent where the gas density in the plateau ($n_{\rm H}^{\rm plateau}$) was determined by molecular observations (cf. references in Table \ref{tab:fit_results}) and $d_{\rm front}$ was evaluated at half of the peak intensity of the mid-IR emission (Figs. \ref{fig:templatesmap} and \ref{fig:allmaps}). $G_0^{\rm front}$ has been evaluated as explained in Sect.\,\ref{pdrsample}. The free parameters of the fit are $n_{\rm H}^0$, $\alpha$ and $d_0$, which define the gas density profile, and the cloud physical length $L$.   The fit results for each one of the PDRs with spatially resolved mid-IR populations are summarised in Table \ref{tab:fit_results}, and the modelled profiles are reported in the middle panels of Fig. \ref{fig:pdrprofiles}. Along with $n_{\rm H}^0$, $\alpha$ and $d_0$, the model enables one to calculate the values of the UV radiation field $G_0$ along the cuts (Fig. \ref{fig:pdrprofiles}). Good fits to $I_0$ are obtained (Fig. \ref{fig:pdrprofiles}), except for the deepest parts of the PDRs, mainly at $d\gtrsim d_0$ where the density is poorly constrained. Since we concentrated our analysis around the AIB peak, this discrepancy does not affect the rest of our analysis.

\end{appendix} 

\begin{acknowledgements}
 We thank the referees  for useful comments that improved the clarity of the paper. This work was supported by the
French National Program, Physique et Chimie du Milieu
Interstellaire, which is gratefully acknowledged.
This paper was also partially supported by the Spanish program CONSOLIDER INGENIO 2010, under grant CSD2009-00038 Molecular Astrophysics: The Herschel and ALMA Era (ASTROMOL).\\
\end{acknowledgements}

\bibliographystyle{aa}
\bibliography{biblio}

\end{document}